\documentclass{article}

\usepackage[letterpaper,top=2cm,bottom=2cm,left=3cm,right=3cm,marginparwidth=1.75cm]{geometry}

\usepackage{amsmath,amssymb}
\usepackage{graphicx}
\usepackage{xcolor}
\usepackage[numbers,sort&compress]{natbib}
\usepackage{multirow} 
\usepackage[colorlinks=true, allcolors=blue]{hyperref}
\usepackage{algorithm,algpseudocode}

\usepackage{booktabs}
\usepackage{caption}
\usepackage{subcaption}
\usepackage{tabularx}
\usepackage{placeins}

\title{Comprehensive Examination of Unrolled Networks for Solving Linear Inverse Problems}
\author{Eric Chen\thanks{Department of Statistics,
Columbia University, NY, USA}\and  Xi Chen\thanks{Department of Electrical and Computer Engineering, Rutgers
University, New Brunswick, NJ, USA} \and Arian Maleki\footnotemark[1] \and Shirin Jalali\footnotemark[2] }

\begin{document}
\maketitle

\begin{abstract}
Unrolled networks have become prevalent in various computer vision and imaging tasks. Although they have demonstrated remarkable efficacy in solving specific computer vision and computational imaging tasks, their adaptation to other applications presents considerable challenges. This is primarily due to the multitude of design decisions that practitioners working on new applications must navigate, each potentially affecting the network's overall performance. These decisions include selecting the optimization algorithm, defining the loss function, and determining the number of convolutional layers, among others. Compounding the issue, evaluating each design choice requires time-consuming simulations to train, fine-tune the neural network, and optimize for its performance. As a result, the process of exploring multiple options and identifying the optimal configuration becomes time-consuming and computationally demanding. The main objectives of this paper are (1) to unify some ideas and methodologies used in unrolled networks to reduce the number of design choices a user has to make, and (2) to report a comprehensive ablation study to discuss the impact of each of the choices involved in designing unrolled networks and present practical recommendations based on our findings. We anticipate that this study will help scientists and engineers design unrolled networks for their applications and diagnose problems within their networks efficiently.
\end{abstract}

\section{Unrolled Networks for Linear Inverse Problems}

In many imaging applications, ranging from magnetic resonance imaging (MRI) and computational tomography (CT scan) to seismic imaging and nuclear magnetic resonance (NMR), the measurement process can be modeled in the following way:
\[
y= Ax^* +w 
\]
In the above equation, $y \in \mathbb{R}^m$ represents the collected measurements and $x^* \in \mathbb{R}^n$ denotes the vectorized image that we aim to capture. The matrix $A \in \mathbb{R}^{m \times n}$ represents the forward operator or measurement matrix of the imaging system, which is typically known exactly or with some small error. Finally, $w$ represents the measurement noise, which is not known, but some information about its statistical properties (such as the approximate shape of the distribution) may be available.

Recovering $x^*$ from the measurement $y$ has been extensively studied, especially in the last 15 years after the emergence of the field of compressed sensing \cite{donoho_compressed_2006, lustig_sparse_2007, candes_introduction_2008, baraniuk2007compressive}. In particular, between 2005-2015, many successful algorithms have been proposed to solve this problem, including Denoising-based AMP \cite{donoho_message-passing_2009, metzler_denoising_2016}, Plug-and-Play Priors \cite{venkatakrishnan_plug-and-play_2013}, compression-based recovery \cite{jalali2016compression, beygi2019efficient}, and Regularization by Denoising \cite{romano_little_2017}. 

Inspired by the successful application of  neural networks, many researchers have started exploring the application of neural networks to solve linear inverse problems \cite{yang_deep_2016, mousavi_learning_2017, chang_one_2017, mousavi_deepcodec_2017, metzler_learned_2017, mccann_convolutional_2017, zhang_ista-net_2018, diamond_unrolled_2018, schlemper_deep_2018, gilton_neumann_2019, aggarwal_modl_2019, ongie_deep_2020, veen_compressed_2020, gilton_deep_2021, gilton_model_2021, NEURIPS2021_6e289439, shastri_denoising_2022, rout_beyond_2023, zhang_physics-inspired_2023, kamilov2023plug, gan2024block, gan2024ptychodv, hu_stochastic_2024, chung_decomposed_2024, chen_practical_2024, chen_self-supervised_2024, shafique_mri_2024, gan_ptychodv_2024}. The original networks proposed for this goal were deep networks that combined convolutional and fully-connected layers \cite{mousavi_learning_2017, kulkarni_reconnet_2016}. The idea was that we feed $y$ or $A^Ty$ to a network and then expect the network to eventually return $x^*$. 

While these methods performed reasonably well in some applications, in many cases, they underperformed the more classical non-neural network based algorithms and simultaneously required computationally demanding training. Some of the challenges that these networks face are as follows: 

\begin{itemize}
\item \textbf{Size of the measurement matrix}: Note that the forward model has matrix $A$ with $m \times n$ elements. Even if we think of small images, we may have $n = 256 \times 256$ and $m = 128 \times 128$. This means that the measurement matrix may have more than $1$ billion elements. Consequently, an effective deep learning-based recovery algorithm may need to memorize the elements or learn the structural properties of $A$ to be able to reconstruct $x^*$ from $y$. This means that the neural network itself should ideally have many more parameters. Not only is the training of such networks computationally very demanding, but also within the computational limits of the work that has been done so far, end-to-end networks have not been very successful.

\item \textbf{Changes in the measurement matrix}: Another issue that is faced by such large networks is that usually one needs to redesign a network and to train a model specific to each measurement matrix. Each network often suffers from poor generalizability to even small changes in the matrix entries. 
\end{itemize}

To address the issues faced by such deep and complex networks in solving inverse problems, and inspired by iterative algorithms for solving convex and non-convex problems, a category of networks known as unrolled networks has emerged \cite{10.5555/3104322.3104374, monga_algorithm_2020}. To understand the motivation behind these unrolled networks, we consider the hypothetical situation in which all images of interest belong to a set $\mathcal{C} \subset \mathbb{R}^n$. Under this assumption, one way to recover the image $x^*$ from the measurements $y$ is to find
\begin{eqnarray}
\arg \min_{x \in \mathcal{C}} \|y-Ax\|_2^2. 
\end{eqnarray}

\begin{figure}
\begin{center}
\includegraphics[width=15cm]{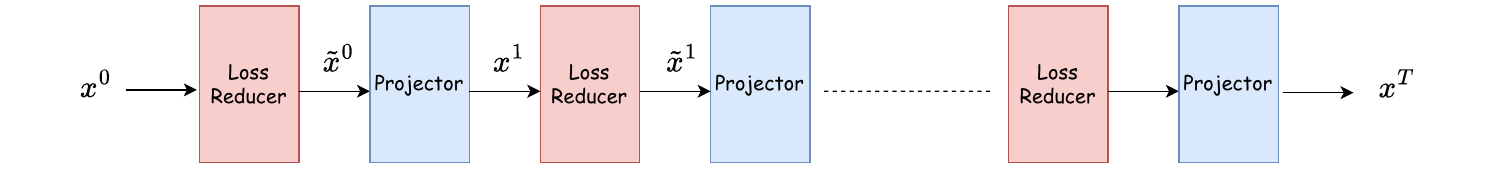}
\caption{Diagram of projected gradient descent. Starting with $x^0 = 0$, the $i^{\rm th}$ loss reducer unit  performs the operation $\tilde{x}^i = x^i + \mu A^T (y-Ax^i)$, and the $i^{\rm th}$ projector unit performs $x^{i+1} = P_{\mathcal{C}} (\tilde{x}^i)$.} \label{fig:unrollednetworkdiagram}
\end{center}
\end{figure}

One method to solve this optimization problem is via projected gradient descent (PGD) that uses the following iterative steps:
\begin{eqnarray}
\label{eq: pgd}
\tilde{x}^i &=& x^i + \mu A^T (y- Ax^i), \nonumber \\
x^{i+1} &=& P_{\mathcal{C}} (\tilde{x}^i). 
\end{eqnarray}

where $x^i$ is the estimate of $x^*$ in iteration $i$, $\mu$ is the step size (learning rate), and $P_{\mathcal{C}}$ denotes the projection onto the set $\mathcal{C}$. Figure \ref{fig:unrollednetworkdiagram} shows the graph of the projected gradient descent algorithm. 

One of the challenges in using PGD for linear inverse problems is that the set $\mathcal{C}$ is unknown, and hence $P_{\mathcal{C}}$ is also not known. For example, $\mathcal{C}$ can represent all natural images of a certain size and $P_{\mathcal{C}} (\cdot)$ a projection onto that space. Researchers have considered ideas such as using state-of-the-art image compression algorithms and image denoising algorithms for the projection step \cite{metzler_denoising_2016, venkatakrishnan_plug-and-play_2013, romano_little_2017, jalali2016compression, beygi2019efficient, rezagah2017compression}, and have more recently adopted neural networks as a promising approach \cite{yang_deep_2016, diamond_unrolled_2018, mardani_neural_2018}. In these formulations, all the projectors in Figure \ref{fig:unrollednetworkdiagram} are replaced with neural networks (usually, the same neural network architecture can be used at different steps). There are several analytical and computational benefits to such an approach: 

\begin{itemize}
\item We do not require a heuristically pre-designed denoiser or compression algorithm to act as the projector $P_{\mathcal{C}}(\cdot)$. Instead, we can deploy a training set to train the networks and optimize their performance. This enables the algorithm to potentially achieve better performance. 

\item Although projected gradient descent analytically employs the same projection operator at every iteration, once we replace them with neural networks, we do not need to impose the constraint of all the neural networks' learned parameters being the same. In fact, giving more freedom can enable us to train the networks more efficiently and, at the same time, improve performance. 

\item Using neural networks enable greater flexibility and can integrate with a wide range of iterative optimization algorithms. For instance, although the above formulation of the unrolled network follows from projected gradient descent, one can also design unrolled networks using a wide range of options, including heavy-ball methods and message passing algorithms.

\end{itemize}

The above formulation of combining projected gradient descent with neural networks belongs to a family called \textit{deep unrolled networks} or \textit{deep unfolded networks}, which is a class of neural network architectures that integrate iterative model-based optimization algorithms with data-driven deep learning approaches \cite{monga_algorithm_2020, li_deep_2021, zhang_physics-inspired_2023}. The central idea is to ``unroll" an iterative optimization algorithm a given number of times, where each iteration replaces the traditional mapping or projection operator with a neural network. The parameters of this network are typically learned end-to-end.  By enabling parameters to be learnable within this framework, unrolled networks combine the interpretability and convergence properties of traditional algorithms with the adaptivity and performance of deep learning models. Due to their efficacy, these networks have been widely used in solving linear inverse problems.

The unrolled network can also be constructed by replacing the projected gradient descent algorithm with various other iterative optimization algorithms. Researchers have explored incorporating deep learning-based projectors with a range of iterative methods, including the Alternating Direction Method of Multipliers (ADMM-Net) \cite{yang_deep_2016}, Iterative Soft-Thresholding Algorithm (ISTA-Net) \cite{zhang_ista-net_2018}, Nesterov's Accelerated First Order Method \cite{beck_fast_2009, zeng_gsista-net_2024}, and Approximate Message Passing (AMP-Net) \cite{zhang_amp-net_2021}. Many of these alternatives offer faster convergence for convex optimization problems, raising the prospect of reducing the number of neural network projectors required, thereby lowering the computational complexity of both training and deploying these networks.

\section{Challenges in Using Unrolled Networks}

As discussed above, the flexibility of unrolled networks has established them as a powerful tool for solving imaging inverse problems. However, applying these architectures to address specific inverse problems presents significant challenges to users. These difficulties stem primarily from two factors: (i) a multitude of design choices and (ii) robustness to noise, measurement matrix, and image resolution. We clarify these issues and present our approach to addressing them below.

\subsection{Design Choices}
The first challenge lies in the numerous design decisions that users must make when employing unrolled networks. We list some main choices below:

\begin{itemize}

\item \textbf{Optimization Algorithm}. In training any unrolled network, the user must decide on which iterative optimization algorithm to unroll. Choices include Projected Gradient Descent, heavy-ball methods (such as Nesterov's accelerated first-order method), Approximate Message Passing (AMP), Alternating Direction Method of Multiplies (ADMM), among others. Unrolled networks trained on different optimization algorithms may lead to drastically varying performances for the task at hand.

\item \textbf{Loss Function}. For any unrolled optimization algorithm, given any observation $y$, one produces a sequence of $T$ projections $\{x^i\}_{i = 1}^{T}$ (See Figure \ref{fig:unrollednetworkdiagram} for an illustration). To train the model, the convention is to define the loss function with respect to the final projection  $x^T$ using the $\ell^2$ loss $\|x^T - x^*\|_2^2$ since this is usually the quantity returned by the network. However, given the non-convexity of the cost function that is used during training, there is no guarantee that the above loss function is optimal for the generalization error. For example, one could use a loss function that incorporates one or more estimates from intermediate stages, such as \( \|x^T - x^*\|_2^2 + \| x^{T/2} - x^*\|_2^2 \), to potentially achieve a better training that provides an improved estimate of \( x^* \) from \( y = Ax^* \). 

As will be shown in our simulation results, the choice of the loss function has a major impact on the performance of the learned networks. Various papers have considered a wide range of loss functions for training different networks. We categorize them broadly below.

\begin{itemize}
\item \textbf{Last Layer Loss}. Consider the notations used in Figure \ref{fig:unrollednetworkdiagram}. The last layer loss evaluates the performance of the network using the following loss function:
\begin{align}
   \ell_{ll}(x^T, x^*) = \|x^T - x^*\|_2^2.
\end{align}
The last layer loss is the most popular loss function that has been used in applications. The main argument for using this loss is that since we only care about the final estimate and that is used as our final reconstruction, we should consider the error of the last estimate. 

\item \textbf{Weighted Intermediate Loss}. While the loss function above seems reasonable, some works in related fields have proposed using an intermediate loss function instead \cite{georgescu_convolutional_2020, mou_deep_2022}. We define the general version of the intermediate loss function as follows:
\begin{align}
\ell_{i,\omega}(x^{1}, x^{2}, \ldots, x^T, x^* )= \sum_{i = 1}^{T}\omega^{T - i}||{x}^i - x^*||_2^2,
\label{equation: intermediate loss}
\end{align}
where $\omega \in (0,1]$. One argument that motivates the use of such loss functions is that, if the predicted image after each projection is closer to the ground truth $x^*$, then it will help the subsequent steps to reach better solutions. The weighted intermediate loss tries to find the right balance among the accuracy of the estimates at different iterations \cite{georgescu_convolutional_2020}. In addition, we make the following observations:

\begin{itemize}
\item When $\omega = 1$, the losses from different layers of the unrolled network are weighted equally. This means that our emphasis on the performance of the last layer is ``weakened." However, this is not necessarily undesirable. As we will show in our simulations, improving the estimates of the intermediate steps will also help improve the recovery quality of $x^T$. 

\item As we decrease the value of $\omega \to 0$, we see that the loss function $\ell_{i, \omega}$ approaches the last layer loss $\ell_{ll}$. The choice of the $\omega$ therefore enables us to interpolate between the two cases. 
\end{itemize}

\item \textbf{Skip $L$-Layer Intermediate Loss}. Another loss function that we investigate is what we call the skip $L$-layer intermediate loss. This loss is similar to the loss used in Inception networks for image classification \cite{szegedy2016rethinking}. Let $L$ be a factor of $T$. Then, the skip $L$-layer loss is given by
\begin{align}
\ell_{s, L} (x^{1}, x^{2}, \ldots, x^T, x^*) = \sum_{i= 0}^{T/L-1} \|x^{T- iL} - x^*\|_2^2. 
\label{equation: skip loss}
\end{align}
For instance, if $T = 15$ and $L = 3$, the skip 3-layer intermediate loss will evaluate the sum of the mean-squared errors between $x^*$ and projections $x^3, x^6, x^9, x^{12},$ and $x^{15}$. By ranging $L$ from $1$ to $T$, one can again interpolate between the two loss function $\ell_{i, 1}$ and $\ell_{l l}$. 

\end{itemize}

As we will show and discuss later in our simulations, it turns out that the unrolled networks that are trained using the above three loss functions exhibit very different performances.

\item \textbf{Number of Unrolled Steps}. Practitioners also have to decide on the number of steps $T$ to unroll for any optimization algorithm. Increasing $T$ often comes with additional computational burdens and may also lead to overfitting. A proper choice of $T$ can ensure that network training is not prohibitively expensive and ensure desirable levels of performance.

\item \textbf{Complexity of the Neural Network}. Similar to the above, the choice of $P_{\mathcal{C}}$ also has a significant impact on the performance of the network. The options entail the number of layers or depth of the network, the activation function to use, whether or not to include residual connections, etc. If the projector is designed to have only little capacity, the unrolled network may have poor recovery. However, if the projector has excessive capacity, the network may become computationally expensive to train and prone to overfitting.

\end{itemize}

It is important to note that after making all the design choices, users are required to conduct time-consuming, computationally demanding, and costly simulations to train the network. Consequently, users may only have the opportunity to explore a limited number of options before settling on their preferred architecture. As a result, in many cases, users are unable to thoroughly evaluate numerous combinations of choices to identify a good network.

\subsection{Robustness and Scaling}

When designing unrolled networks for inverse problems, it is common to aim for robustness across a range of settings beyond the specific conditions for which the algorithm was originally designed. While an algorithm may be tailored for a particular signal type, image resolution, number of observations, or noise level, it is desirable for the network to maintain effectiveness across different settings as well.  

As a simple motivating example, consider when a new imaging device has been acquired that operates using a different observation matrix. If the unrolled network previously designed has bad adaptivity and performance with respect to the current matrix, one would be required to go over the entire process to determine a new batch of choices for the current setting. Therefore, ideally, one would like to have a single network structure that works well across a wide range of applications.

\subsection{Our Approach for Designing Unrolled Networks}

As discussed above, one must face an abundance of design choices before training and deploying an unrolled network. However, testing the performance of all possible enumerations of these choices across a wide range of applications and datasets is computationally demanding and combinatorially prohibitive. This complexity hinders practitioners from applying the optimal unrolled network in their problem-specific applications. To offer a more systematic way for designing such networks, we adopt the following high-level approach:

\begin{itemize}

\item We present the \textbf{Deep Memory Unrolled Network (DeMUN)} where each step of the network leverages the gradient from all previous iterations. These networks encompass various existing models, including those based on gradient descent, Nesterov's method, Approximate Message Passing, and more as special cases. This network lets the data decide on the optimal choice of algorithm to be unrolled, and hence improves recovery performance. A byproduct of our choice is that users do not need to decide on the choice of the algorithm they need to unroll. 

\item We present several hypotheses regarding important design choices that underlie the design of unrolled networks, and we test them using extensive simulations. These hypotheses allow users to avoid exploring the multitude of design choices that they have to face in practice.  

\end{itemize}

These two steps allow users to bypass many design choices, such as selecting an optimization algorithm or loss function, thus simplifying the process of designing unrolled networks. We test the robustness of our hypotheses with respect to the changes in the measurement matrices and noise in the system. These robustness results suggest that the simplified design approach presented in this paper can be applied to a much wider range of systems than those specifically studied here.

\section{Deep Memory Unrolled Network (DeMUN)}

As discussed previously, one of the initial decisions users face when designing an unrolled network is selecting the optimization algorithm to unroll. Various optimization algorithms, including gradient descent, heavy-ball methods, and Approximate Message Passing have been incorporated into unrolled networks. We introduce the Deep Memory Unrolled Network (DeMUN), which encompasses many of these algorithms as special cases. At the $i$-th iteration in DeMUN, the update of $\tilde{x}^i$ is given by:
\begin{equation}\label{eq:LongMemory}
\tilde{x}^{i}= \alpha^i x^i + \sum_{j=0}^i \beta^i_j A^T (y-Ax^{j}). 
\end{equation}
for $i \in \{0, \dots, T - 1\}$ where $x^0 = 0$. In other words, while calculating $\tilde{x}^i$, it uses not only the gradient calculated at the current step, but also leverages all the gradients calculated from previous steps. 

\begin{figure}
\begin{center}
\includegraphics[width= 15cm]{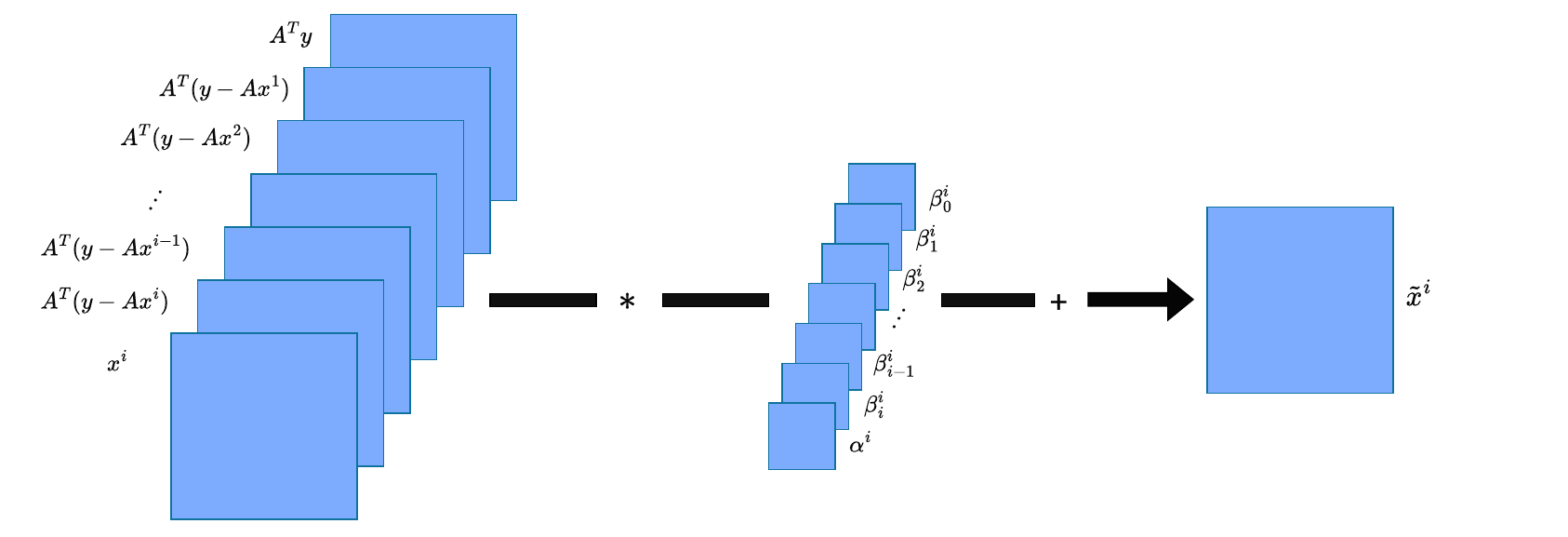}
\caption{An example of the memory terms combined into a single image.} \label{fig:longmemory}
\end{center}
\end{figure}

By using different choices for $\beta_0^i, \beta_1^i, \ldots, \beta_i^i$ at each iteration, one can recover a large class of algorithms including gradient descent, heavy-ball methods, and Approximate Message Passing. As shown in equation \eqref{eq:LongMemory} and illustrated in Figure \ref{fig:longmemory}, we can rearrange the vector $x^i$ and the gradients $\{A^T (y-Ax^{j})\}_{j \in \{0, \dots, i\}}$ as images and view the expression as one-by-one convolutions over the images. Our simulation results reported later show that DeMUNs with trainable $\beta_0^i, \beta_1^i, \ldots, \beta_i^i$ offer greater flexibility and better performance compared to fix instances such as gradient descent or Nesterov's method. 

In the following sections, we present extensive simulation results to: (i) demonstrate that DeMUNs outperform a variety of other unrolled networks that are based on different iterative optimization algorithms, and (ii) explore the optimal design choices users face and their impact on the performance of DeMUNs, guided by a series of hypotheses we propose.

\section{Our Four Main Hypotheses}
\label{section: three hypothesis}

\subsection{Simulation Setup}
Our goal in this section is to 1) show the efficacy of DeMUNs by comparing their performance against different unrolled algorithms, and 2) explore the impact of specific design choices. We conduct extensive ablation studies where we fix all but one design choice at each step and explore the performance of unrolled algorithms under different options for this choice. Based on these studies, we have developed several hypotheses aimed at simplifying the design of unrolled networks. We will outline these hypotheses and present simulation results that support them.

For all simulations below, we report results of four different sampling rates $m/n$ for the measurement matrix $A$: 10\%, 20\%, 30\%, and 40\%. In the initial phase of our simulations, each entry in the measurement matrix is i.i.d.~Gaussian, where $A_{ij} \sim \mathcal{N}(0, 1/m)$ for $A \in \mathbb{R}^{m \times n}$. While we will discuss the impact of the resolution on the performance of the algorithms, in the initial simulations, all training images have resolution $50 \times 50$ and vectorizing the images leads to $n = 2500$. In addition, we consider three different settings for the number of unrolled steps $T$: 5, 15, and 30. For all results below, we report the Peak Signal-to-Noise Ratio (PSNR) for the networks trained under the aforementioned sampling rates and number of projection steps on a test set of 2500 images. More details on data collection and processing, training of unrolled networks, and evaluation are deferred to Appendix \ref{Experimental Setup}.

In our simulations, we adopt the general DnCNN architecture as outlined by Zhang et al. as our neural network projector $P_{\mathcal{C}}$ \cite{zhang_beyond_2017}. A DnCNN architecture with $L$ intermediate layers consists of an input layer with 64 filters of size $3 \times 3 \times 1$ followed by a ReLU activation function to map the input image to 64 channels\footnote{It is of size $3 \times 3 \times 1$ since we assume the images are in grayscale.}, $L$ layers consisting of 64 filters of size $3 \times 3 \times 64$ followed by BatchNormalization and ReLU, and a final reconstruction layer with a single filter of size $3 \times 3 \times 64$ to map to the output dimension of $50 \times 50 \times 1$. It should be noted that this architecture is very flexible in general as one is not constrained to using 64 channels or a fixed number of layers, but has the freedom to vary parameters of the network based on the problem at hand. We start with using $L = 5$ intermediate layers without further specification. However, we will also study the impact of the number of layers in DnCNN in Sections \ref{ssec:complexityimpact} and \ref{subsection: projector capacity}.

\subsection{Overview of Our Simplifying Hypotheses}

As previously described, we begin with four hypotheses, each of which contributes to improving the performance of unrolled networks, enhancing training practices, and simplifying the design process by reducing the number of decisions practitioners need to make. These hypotheses are based on extensive simulations, some of which are reported below.
\bigbreak
\noindent \textbf{Hypothesis 1:} \textit{ Unrolled networks trained with the loss function $\ell_{i, 1}$ uniformly outperform their counterparts trained with $\ell_{ll}$. Among the unrolling algorithms we tested, i.e., PGD, AMP, and Nesterov, DeMUNs offer superior recovery performance.} \\

\noindent Although we are primarily concerned with the quality of the final reconstruction $\|x^T -x^*\|_2^2$, we find that using the loss function $\ell_{i,1} = \sum_{i = 1}^{T}||x^i - x^*||_2^2$ during training yields better recovery performance than focusing solely on the last layer. This improvement may be attributed to the smoother optimization landscape provided by using the intermediate loss, which guides the network more effectively towards better minima. We present our empirical evidence for suggesting this hypothesis in Section \ref{subsection: Hypothesis 1}. With the advantage of using an unweighted intermediate loss function established, we next explore the impact of incorporating residual connections into unrolled networks. \\

\noindent \textbf{Hypothesis 2:} \textit{DeMUNs trained using residual connections and loss function $\ell_{i,1}$ uniformly improve recovery performance compared to those trained without residual connections.}\\

\noindent Residual connections are known to alleviate issues such as vanishing gradients and facilitate the training of deeper networks by allowing gradients to propagate more effectively through the intermediate layers \cite{he_deep_2015, he_identity_2016}. Specifically, we modify each projection step in our unrolled network to be $x^{i + 1} = \tilde{x}^i + P_{\mathcal{C}}(\tilde{x}^i)$. In verifying Hypothesis 2, we continue to use $\omega = 1$ (see the definition of intermediate loss in \eqref{equation: intermediate loss}). This ensures that any observed improvements can be directly attributed to the addition of residual connections rather than changes in the loss function. We present our empirical evidence for suggesting this hypothesis in Section \ref{subsection: Hypothesis 2}. Having confirmed that both the use of an unweighted intermediate loss and the inclusion of residual connections improve recovery performance, we further investigate the sensitivity of our network to the specific shape of the loss function. \\

\noindent \textbf{Hypothesis 3:} \textit{For training DeMUNs, there is no significant difference among the following loss functions: (1) $\ell_{i, 1}$, (2) $\ell_{i, 0.95}$, (3) $\ell_{i, 0.85}$. Furthermore, $\ell_{i, 0.5}$, $\ell_{i, 0.25}$, $\ell_{i, 0.1}$, $\ell_{i, 0.01}$, and $\ell_{s,5}$ perform worse than $\ell_{i, 1}$. }  \\ 

\noindent We present our empirical evidence for suggesting this hypothesis in Section \ref{subsection: hypothesis 3}. \\

\noindent \textbf{Hypothesis 4:} \textit{When we vary the number of layers, $L$, in the DnCNN from $5$ to $15$, the performance of DeMUNs remains largely unchanged, indicating that the number of layers has a negligible impact on its performance. However, increasing $L$ from \(3\) to \(5\) provides a noticeable improvement in performance. } \\ 

\noindent We present our empirical evidence for this hypothesis in Section \ref{ssec:complexityimpact}. 
Confirming these hypotheses provides a set of practical recommendations for designing unrolled networks that are both effective and robust across various settings. In the following four sections, we summarize some of the simulations that have led to the above hypotheses. 

\begin{figure}
\begin{center}
\includegraphics[width= 15cm]{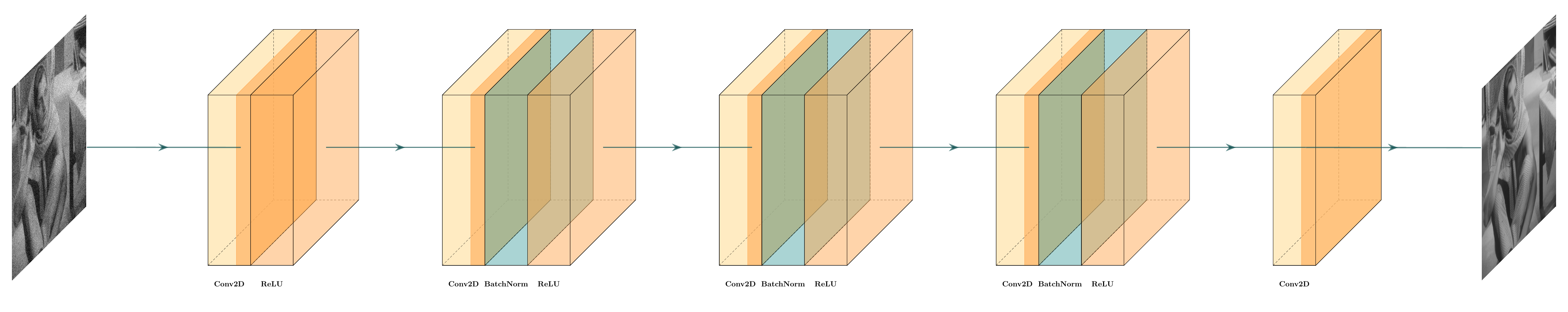}
\caption{An example of the DnCNN architecture with $L = 3$ intermediate layers} \label{fig:DnCNN}
\end{center}
\end{figure}

\subsection{Impact of Intermediate Loss}
\label{subsection: Hypothesis 1}

In this section, we aim to validate Hypothesis 1, which posits that deep unrolled networks trained with the unweighted intermediate loss function $\ell_{i, 1}$ uniformly outperform their counterparts trained with the last layer loss $\ell_{ll}$. Among the various algorithms, we hypothesize that the Deep Memory Unrolled Network offers superior recovery performance compared to other unrolled algorithms. In the following, we consider four different unrolled algorithms:\footnote{See Appendix \ref{appendix: unrolled algs} for details of Nesterov and AMP. }

\begin{enumerate}
    \item Deep Memory Unrolled Network (DeMUN): Our proposed network that incorporates the memory of all the gradients into the unrolling process.
    \item Projected Gradient Descent (PGD): The standard unrolled algorithm outlined in \eqref{eq: pgd}.
    \item Nesterov's Accelerated First-Order Method (Nesterov): An optimization method that uses momentum to accelerate convergence \cite{beck_fast_2009}.
    \item Approximate Message Passing (AMP): An iterative algorithm tailored for linear inverse problems with Gaussian sensing matrices \cite{metzler_denoising_2016, metzler_learned_2017}. 
\end{enumerate}

For all unrolled algorithms, we are considering when all the projection steps are cast as direct projections of the form $x^{i + 1} = P_{\mathcal{C}}(\tilde{x}^i)$ and compare the performance between last layer loss and unweighted intermediate loss. 

\begin{table}[ht!]
\small
    \begin{subtable}[ht]{0.45\textwidth}
        \centering
        \begin{tabular}{||c c c c c||} 
         \hline
         $m$ & DeMUN & PGD & Nesterov & AMP \\ [0.5ex] 
         \hline\hline
         $0.1n$ & 25.22 & 25.10 & 24.55 & 15.20\\ 
         \hline
         $0.2n$ & 28.03 & 27.42 & 27.68 & 21.25\\ 
         \hline
         $0.3n$ & 29.53 & 29.53 & 28.67 & 25.00\\ 
         \hline
         $0.4n$ & 31.70 & 30.85 & 30.43 & 24.07\\ 
         \hline
        \end{tabular}
       \caption*{Last Layer Loss $\ell_{ll}$}
    \end{subtable}
    \hfill
    \begin{subtable}[ht]{0.45\textwidth}
        \centering
        \begin{tabular}{||c c c c c||} 
         \hline
         $m$ & DeMUN & PGD & Nesterov & AMP \\ [0.5ex] 
         \hline\hline
         $0.1n$ & 26.09& 25.80& 25.74 & 25.11\\ 
         \hline
         $0.2n$ & 28.70& 28.04& 28.10 & 27.50\\ 
         \hline
         $0.3n$ & 30.60& 29.78& 29.50 & 29.19\\ 
         \hline
         $0.4n$ & 31.87& 31.33& 30.60 & 30.70\\ 
         \hline
        \end{tabular}
        \caption*{Intermediate Loss $\ell_{i, 1}$} 
     \end{subtable}
     \caption{Average PSNR (dB) for 2500 Test Images Under \textbf{5} Projection Steps}
     \label{tab:LL_IL_P5}
\end{table}

\begin{table}[ht!]
\small
    \begin{subtable}[ht]{0.45\textwidth}
        \centering
        \begin{tabular}{||c c c c c||} 
         \hline
         $m$ & DeMUN & PGD & Nesterov & AMP \\ [0.5ex] 
         \hline\hline
         $0.1n$ & 24.42 & 24.72 & 24.56 & 19.94\\ 
         \hline
         $0.2n$ & 27.23 & 27.71 & 26.36 & 23.43\\ 
         \hline
         $0.3n$ & 24.78 & 28.99 & 28.41 & 23.35\\ 
         \hline
         $0.4n$ & 30.20 & 30.53 & 26.47 & 24.71\\ 
         \hline
        \end{tabular}
       \caption*{Last Layer Loss $\ell_{ll}$}
    \end{subtable}
    \hfill
    \begin{subtable}[ht]{0.45\textwidth}
        \centering
        \begin{tabular}{||c c c c c||} 
         \hline
         $m$ & DeMUN & PGD & Nesterov & AMP \\ [0.5ex] 
         \hline\hline
         $0.1n$ & 26.73 & 26.42& 26.07 & 26.57\\ 
         \hline
         $0.2n$ & 29.97 & 28.97& 28.39 & 29.11\\ 
         \hline
         $0.3n$ & 31.89 & 30.95& 30.18 & 30.99\\ 
         \hline
         $0.4n$ & 33.68 & 32.52& 31.47 & 32.72\\ 
         \hline
        \end{tabular}
        \caption*{Intermediate Loss $\ell_{i, 1}$}
     \end{subtable}
     \caption{Average PSNR (dB) for 2500 Test Images Under \textbf{15} Projection Steps}
     \label{tab:LL_IL_P15}
\end{table}

\begin{table}[ht!]
\small
    \begin{subtable}[ht]{0.45\textwidth}
        \centering
        \begin{tabular}{||c c c c c||} 
         \hline
         $m$ & DeMUN & PGD & Nesterov & AMP \\ [0.5ex] 
         \hline\hline
         $0.1n$ & 25.27 & 24.17 & 24.49 & 16.04\\ 
         \hline
         $0.2n$ & 26.37 & 27.30 & 27.00 & 19.99\\ 
         \hline
         $0.3n$ & 28.44 & 28.30 & 29.24 & 23.21\\ 
         \hline
         $0.4n$ & 31.31 & 30.33 & 29.95 & 22.66\\ 
         \hline
        \end{tabular}
       \caption*{Last Layer Loss $\ell_{ll}$}
    \end{subtable}
    \hfill
    \begin{subtable}[ht]{0.45\textwidth}
        \centering
        \begin{tabular}{||c c c c c||} 
         \hline
         $m$ & DeMUN & PGD & Nesterov & AMP \\ [0.5ex] 
         \hline\hline
         $0.1n$ & 26.96 & 26.51& 26.19 & 26.73\\ 
         \hline
         $0.2n$ & 29.86 & 29.07& 28.35 & 29.12\\ 
         \hline
         $0.3n$ & 32.05 & 30.88& 29.56 & 31.24\\ 
         \hline
         $0.4n$ & 34.05 & 32.33& 31.15 & 32.87\\ 
         \hline
        \end{tabular}
        \caption*{Intermediate Loss $\ell_{i, 1}$}
     \end{subtable}
     \caption{Average PSNR (dB) for 2500 Test Images Under \textbf{30} Projection Steps}
     \label{tab:LL_IL_P30}
\end{table}

\begin{figure}
\begin{center}
\includegraphics[width= 1.0\textwidth]{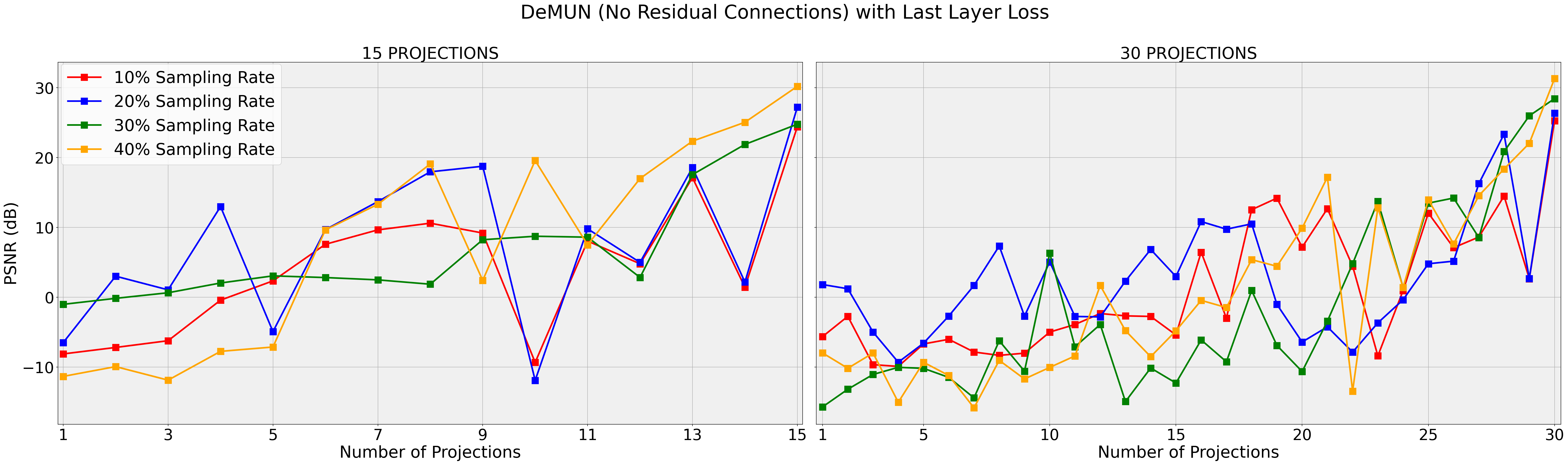}
\caption{DeMUN (no residual connections) with loss $\ell_{ll}$. The networks are trained for $T=15$ (left) and $T=30$ (right), and the graph displays the PSNR after each intermediate projection.}
\label{fig:DeMUN_NR_LL}
\end{center}
\end{figure}

\begin{figure}
\begin{center}
\includegraphics[width= 1.0\textwidth]{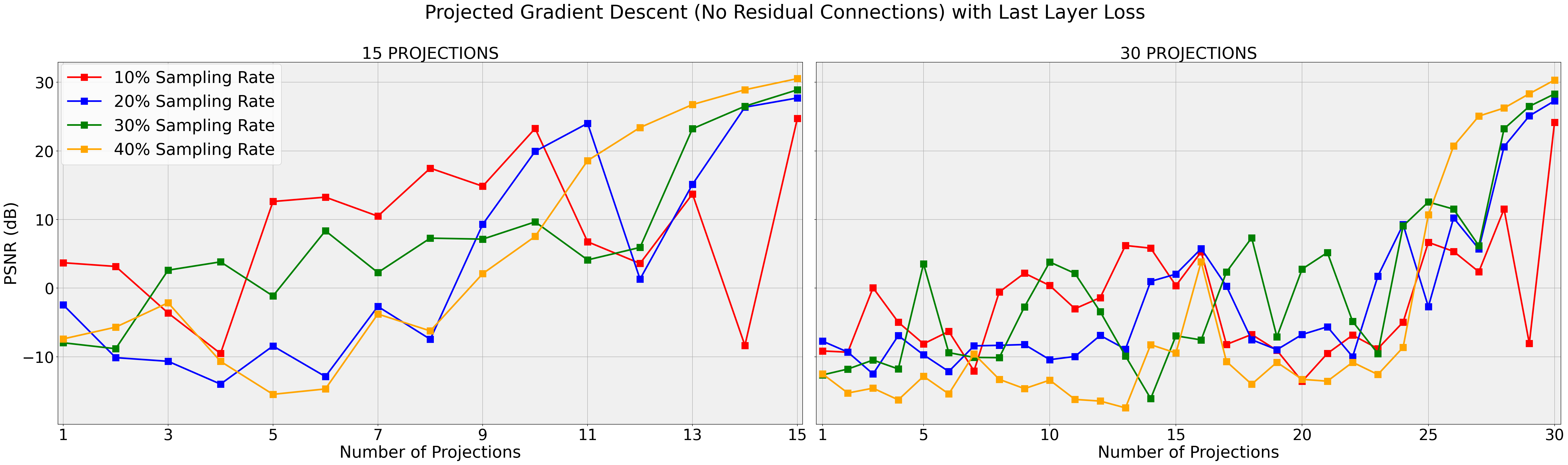}
\caption{PGD (no residual connections) with loss $\ell_{ll}$. The networks are trained for $T=15$ (left) and $T=30$ (right), and the graph displays the PSNR after each intermediate projection.}
\label{fig:PGD_NR_LL}
\end{center}
\end{figure}

\begin{figure}
\begin{center}
\includegraphics[width= 1.0\textwidth]{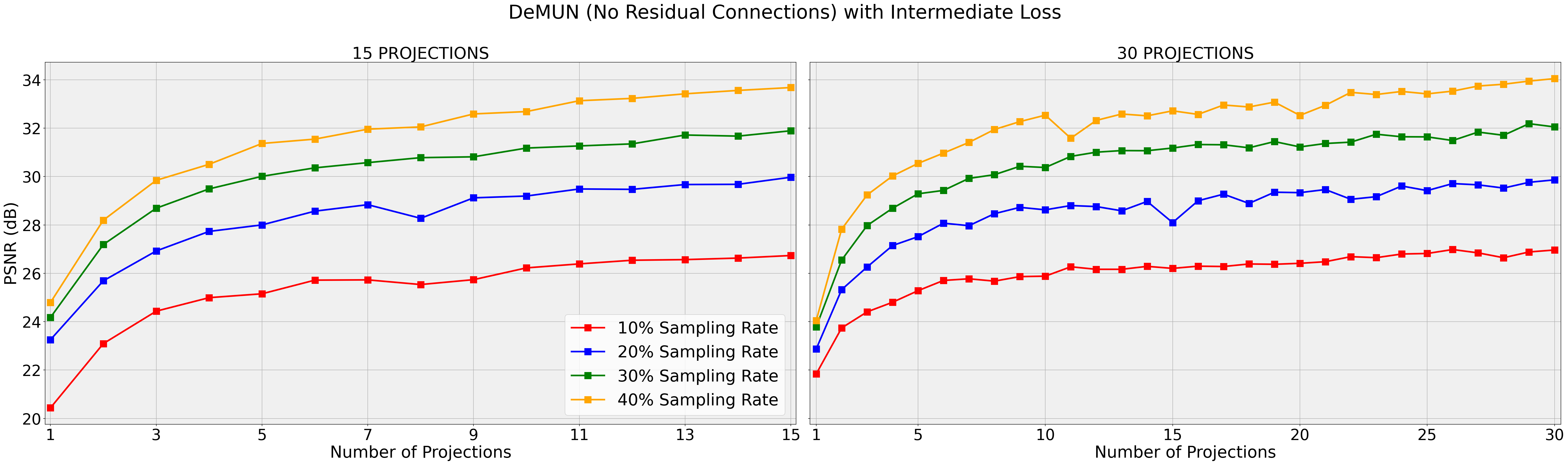}
\caption{DeMUN (no residual connections) with loss $\ell_{i,1}$. The networks are trained for $T=15$ (left) and $T=30$ (right), and the graph displays the PSNR after each intermediate projection.}
\label{fig:DeMUN_NR_IL}
\end{center}
\end{figure}

\begin{figure}
\begin{center}
\includegraphics[width= 1.0\textwidth]{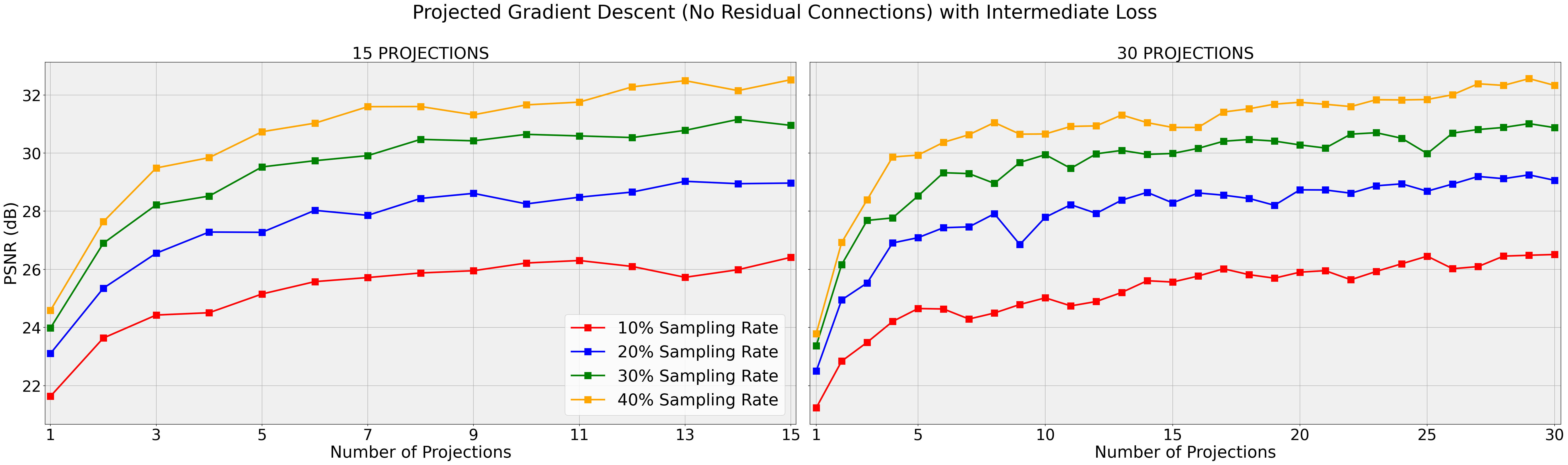}
\caption{PGD (no residual connections) with loss $\ell_{i,1}$. The networks are trained for $T=15$ (left) and $T=30$ (right), and the graph displays the PSNR after each intermediate projection.}
\label{fig:PGD_NR_IL}
\end{center}
\end{figure}

\FloatBarrier

From the above results and plots of the intermediate PSNR recovery, we make the following remarks.

\begin{itemize}
\item \textbf{Improved Performance with Intermediate Loss:} 

By analyzing the tables and graphs, we conclude that across all four unrolled algorithms, training with the intermediate loss function $\ell_{i, 1}$, consistently yields higher PSNR values compared to training with the last layer loss, $\ell_{ll}$. In particular, the improvement offered by the intermediate loss function becomes more significant as the number of projections increases.

To illustrate this point, consider the performance of DeMUN at an undersampling rate of $0.4n$. When training with the last layer loss $\ell_{ll}$, the PSNR values are $31.7$dB, $30.2$dB, and $31.31$dB corresponding to $5$ (Table \ref{tab:LL_IL_P5}), $15$ (Table \ref{tab:LL_IL_P15}), and $30$ (Table \ref{tab:LL_IL_P30}) projections respectively. These results indicate that increasing the number of projections offers no substantial gain. In contrast, when training with the intermediate loss, the performance improves significantly: from $31.87$dB for 5 projections, to $33.68$dB for 15 projections, and further to $34.05$dB for 30 projections.

Figures \ref{fig:DeMUN_NR_LL} and \ref{fig:DeMUN_NR_IL} help clarify this difference. In these two figures, the networks are trained for $T = 15$ (left) and $T = 30$ (right) projections, and the figures display the PSNR after each intermediate projection. As shown in Figure \ref{fig:DeMUN_NR_LL}, with 30 projections and the last layer loss, the network's estimates for the first twenty projections are only marginally better than the estimate after the first projection. This stagnation contrasts sharply with the intermediate loss case depicted in Figure \ref{fig:DeMUN_NR_IL}, where each additional projection contributes to improved performance.

In summary, training with the last layer loss fails to leverage the full complexity of the network, whereas the intermediate loss function enables the training process to utilize the network's capacity effectively, resulting in progressively better performance with more projections.

\item \textbf{Superiority of Deep Memory Unrolled Network:} Among all algorithms that we have unrolled, i.e., PGD, Nesterov, and AMP, DeMUN achieves the highest PSNR values when trained with the intermediate loss, confirming our hypothesis.\footnote{However, we see that this is not always the case when using last layer loss. A possible explanation is that our memory networks contain many parameters (especially with many projection steps) and may be stuck at a local minimum during training using the last-layer loss. In contrast, when adopting the intermediate loss function, the network needs to optimize for its projection performance across all projection steps to minimize the loss. As a result, it may find better solutions especially for the parameters that are involved in the earlier layers.} This is to be expected, as DeMUNs encompass the other unrolled networks as special cases. During training, the data effectively determines which algorithm should be unrolled, intuitively speaking.

\item \textbf{No PSNR Decrease in Terms of the Number of Projections with Intermediate Loss:} 
As illustrated in Figures \ref{fig:DeMUN_NR_IL} and \ref{fig:PGD_NR_IL}, when the networks are trained with the intermediate loss, the performance consistently improves as the number of projections increases, making it a non-decreasing function of the number of projections selected. However, these figures also show that the performance begins to plateau after a certain number of projection steps. This property simplifies the design process of unrolled networks.

\end{itemize}

Given the above observations, we conjecture that the intermediate loss may provide several benefits:

\begin{itemize}
    \item \textbf{Avoiding Poor Local Minima:} Focusing solely on the output of the final layer may lead the network to suboptimal solutions (due to nonconvexity). In comparison, the intermediate loss encourages the network to make meaningful progress at each step, which potentially reduces the risk of getting stuck in poor local minima.
    \item \textbf{More Information during Backpropagation:} By including losses from all intermediate steps, the network receives more gradient information during autodifferentiation, which may be helpful in learning better representations and weights.
\end{itemize}

These empirical results strongly support our first hypothesis that incorporating information from all intermediate steps creates a more effective learning mechanism for the network.

\FloatBarrier
\subsection{Impact of Residual Connections}
\label{subsection: Hypothesis 2}

Having verified that training with the intermediate loss function $\ell_{i, 1}$ improves the recovery performance of unrolled networks, we now examine the effect of including residual connections of the form $x^{i + 1} = \tilde{x}^i + P_{\mathcal{C}}(\tilde{x}^i)$ into unrolled networks as stated in hypothesis 2 when fixing the choice of the unweighted intermediate loss function $\ell_{i,1}$. For comparison, in addition to the Deep Memory Unrolled Network, we also include the results for unrolled networks based on PGD under the same conditions. 

\begin{table}[ht!]
\small
\begin{center}
\begin{tabular}{||c c c c c||} 
 \hline
 $m$ & DeMUN (No Residual) & DeMUN (Residual) & PGD (No Residual) & PGD (Residual) \\ [0.5ex] 
 \hline\hline
 $0.1n$ & 26.09 & 26.29 & 25.80 & 26.10 \\ 
 \hline
 $0.2n$ & 28.70 & 29.17 & 28.04 & 28.66 \\ 
 \hline
 $0.3n$ & 30.60 & 30.95 & 29.78 & 30.31 \\ 
 \hline
 $0.4n$ & 31.87 & 32.52 & 31.33 & 31.64 \\ 
 \hline
\end{tabular}
\caption{Average PSNR (dB) for 2500 Test Images Under \textbf{5} Projection Steps}
\end{center}
\end{table}

\begin{table}[ht!]
\small
\begin{center}
\begin{tabular}{||c c c c c||} 
 \hline
 $m$ & DeMUN (No Residual) & DeMUN (Residual) & PGD (No Residual) & PGD (Residual) \\ [0.5ex] 
 \hline\hline
 $0.1n$ & 26.73 & 27.22 & 26.42 & 26.43 \\ 
 \hline
 $0.2n$ & 29.97 & 30.33 & 28.97 & 29.74 \\ 
 \hline
 $0.3n$ & 31.89 & 32.70 & 30.95 & 31.60 \\ 
 \hline
 $0.4n$ & 33.68 & 34.43 & 32.52 & 33.41 \\ 
 \hline
\end{tabular}
\caption{Average PSNR (dB) for 2500 Test Images Under \textbf{15} Projection Steps}
\end{center}
\end{table}

\begin{table}[ht!]
\small
\begin{center}
\begin{tabular}{||c c c c c||} 
 \hline
 $m$ & DeMUN (No Residual) & DeMUN (Residual) & PGD (No Residual) & PGD (Residual) \\ [0.5ex] 
 \hline\hline
 $0.1n$ & 26.96 & 27.44 & 26.51 & 26.61 \\ 
 \hline
 $0.2n$ & 29.86 & 30.74 & 29.07 & 30.07 \\ 
 \hline
 $0.3n$ & 32.05 & 32.77 & 30.88 & 31.89 \\ 
 \hline
 $0.4n$ & 34.05 & 34.86 & 32.33 & 33.74 \\ 
 \hline
\end{tabular}
\caption{Average PSNR (dB) for 2500 Test Images Under \textbf{30} Projection Steps}
\end{center}
\end{table}

\begin{figure}[ht!]
\begin{center}
\includegraphics[width = 1.0\textwidth]{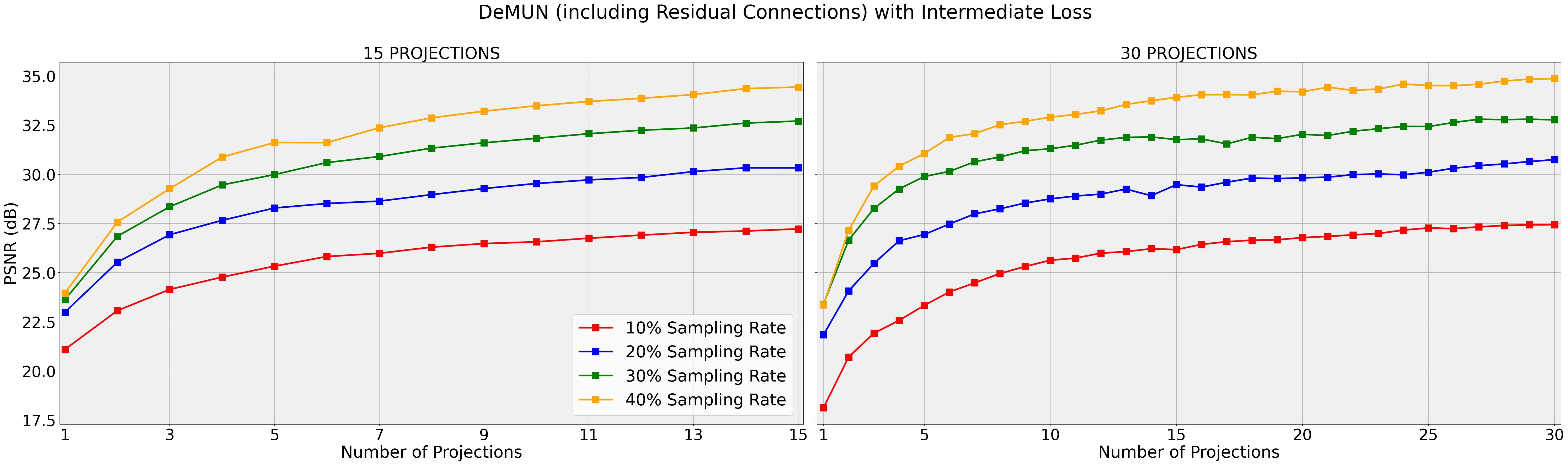}
\caption{DeMUN (including residual connections) with loss $\ell_{i,1}$. The networks are trained for $T=15$ (left) and $T=30$ (right), and the graph displays the PSNR after each intermediate projection.}
\label{fig:DeMUN_R_IL}
\end{center}
\end{figure}

\begin{figure}[ht!]
\begin{center}
\includegraphics[width= 1.0\textwidth]{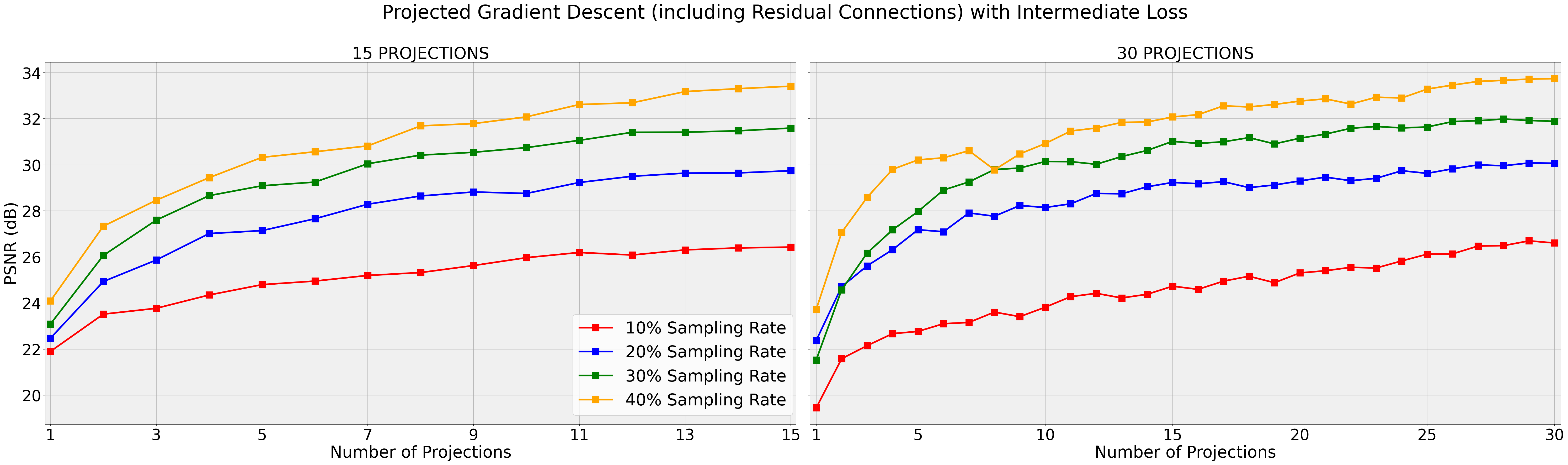}
\caption{PGD (including residual connections) with loss $\ell_{i,1}$. The networks are trained for $T=15$ (left) and $T=30$ (right), and the graph displays the PSNR after each intermediate projection.}
\label{fig:PGD_R_IL}
\end{center}
\end{figure}

\FloatBarrier

From the table and figures above, we observe the following:
\begin{enumerate}
    \item \textbf{Consistent Performance Improvement:} Including residual connections consistently improves the PSNR across all sampling rates and number of projection steps for both the Deep Memory and PGD based unrolled networks.
    \item \textbf{Superior Performance of Deep Memory Network:} While both networks benefit from residual connections, the Deep Memory Unrolled Networks maintain superior performance over Projected Gradient Descent in all scenarios.
\end{enumerate}

These empirical results strongly support Hypothesis 2 that incorporating residual connections into the Deep Memory Unrolled Network further improves its performance on top of training with the unweighted intermediate loss function. The consistent improvement across different sampling rates and projection steps potentially highlights the value of residual connections in unrolled network architectures.

\FloatBarrier
\subsection{Sensitivity to Other Loss Functions}
\label{subsection: hypothesis 3}

Having identified that using an unweighted intermediate loss function and incorporating residual connections in Deep Memory Unrolled Networks offer superior performance, we now explore the sensitivity of our network to variations in the loss function as raised in Hypothesis 3. Specifically, we want to see whether different weighting schemes in the intermediate loss function or using a skip-$L$ layer loss significantly impact the recovery performance. We consider the following variations of the loss function: $\ell_{i, \omega}$ where $\omega \in \{0.95, 0.85, 0.75, 0.5, 0.25, 0.1, 0.01\}$ and $\ell_{s, 5}$. We also include results for Deep Memory Unrolled Networks trained using $\ell_{ll}$ with residual connections for comparison.

\begin{table}[ht!]
\begin{center}
\begin{tabular}{||c c c c c c c c c c c||} 
 \hline
 $m$ & $\ell_{i, 1}$ & $\ell_{i, 0.95}$ & $\ell_{i, 0.85}$ & $\ell_{i, 0.75}$ & $\ell_{i, 0.5}$ & $\ell_{i, 0.25}$ & $\ell_{i, 0.1}$ & $\ell_{i, 0.01}$ & $\ell_{s, 5}$ & $\ell_{ll}$\\ [0.5ex] 
 \hline\hline
 $0.1n$ & 26.29 & 26.35 & 26.34 & 26.20 & 26.00 & 25.76 & 25.43 & 25.53 & 25.33 & 25.32\\ 
 \hline
 $0.2n$ & 29.17 & 29.09 & 29.04 & 29.07 & 28.74 & 28.66 & 28.36 & 28.34 & 28.45 & 28.38\\
 \hline
 $0.3n$ & 30.95 & 30.94 & 30.97 & 30.97 & 30.79 & 30.51 & 30.43 & 30.34 & 30.18 & 30.59\\
 \hline
 $0.4n$ & 32.52 & 32.54 & 32.54 & 32.47 & 32.31 & 32.07 & 31.89 & 31.84 & 31.86 & 31.97\\
 \hline
\end{tabular}
\caption{Average PSNR (dB) for 2500 Test Images Under \textbf{5} Projection Steps} \label{tb:diffloss:proj5}
\end{center}
\end{table}

\begin{table}[ht!]
\begin{center}
\begin{tabular}{||c c c c c c c c c c c||} 
 \hline
 $m$ & $\ell_{i, 1}$ & $\ell_{i, 0.95}$ & $\ell_{i, 0.85}$ & $\ell_{i, 0.75}$ & $\ell_{i, 0.5}$ & $\ell_{i, 0.25}$ & $\ell_{i, 0.1}$ & $\ell_{i, 0.01}$ & $\ell_{s, 5}$ & $\ell_{ll}$\\ [0.5ex] 
 \hline\hline
 $0.1n$ & 27.22 & 26.98 & 27.02 & 26.96 & 26.93 & 26.79 & 26.73 & 26.36 & 27.04 & 26.04\\ 
 \hline
 $0.2n$ & 30.33 & 30.08 & 30.33 & 30.27 & 30.01 & 29.77 & 29.19 & 29.68 & 30.15 & 29.46\\
 \hline
 $0.3n$ & 32.70 & 32.56 & 32.49 & 32.35 & 32.18 & 31.97 & 30.87 & 31.12 & 32.33 & 31.52\\
 \hline
 $0.4n$ & 34.43 & 34.30 & 34.33 & 34.14 & 34.00 & 33.35 & 33.51 & 33.17 & 33.97 & 33.09\\
 \hline
\end{tabular}
\caption{Average PSNR (dB) for 2500 Test Images Under \textbf{15} Projection Steps}\label{tb:diffloss:proj15}
\end{center}
\end{table}

\begin{table}[ht!]
\begin{center}
\begin{tabular}{||c c c c c c c c c c c||} 
 \hline
 $m$ & $\ell_{i, 1}$ & $\ell_{i, 0.95}$ & $\ell_{i, 0.85}$ & $\ell_{i, 0.75}$ & $\ell_{i, 0.5}$ & $\ell_{i, 0.25}$ & $\ell_{i, 0.1}$ & $\ell_{i, 0.01}$ & $\ell_{s, 5}$ & $\ell_{ll}$\\ [0.5ex] 
 \hline\hline
 $0.1n$ & 27.44 & 27.41 & 27.42 & 27.22 & 27.04 & 26.99 & 26.81 & 26.65 & 27.31 & 26.63\\ 
 \hline
 $0.2n$ & 30.74 & 30.63 & 30.46 & 30.62 & 30.07 & 29.92 & 29.28 & 29.63 & 30.52 & 29.39\\
 \hline
 $0.3n$ & 32.76 & 32.92 & 32.79 & 32.58 & 32.22 & 31.98 & 31.79 & 31.66 & 32.72 & 31.65\\
 \hline
 $0.4n$ & 34.86 & 34.36 & 34.79 & 34.55 & 33.74 & 33.46 & 33.03 & 33.50 & 34.72 & 32.89\\
 \hline
\end{tabular}
\caption{Average PSNR (dB) for 2500 Test Images Under \textbf{30} Projection Steps}\label{tb:diffloss:proj30}
\end{center}
\end{table}

\begin{figure}[ht!]
\begin{center}
\includegraphics[width= 1.0\textwidth]{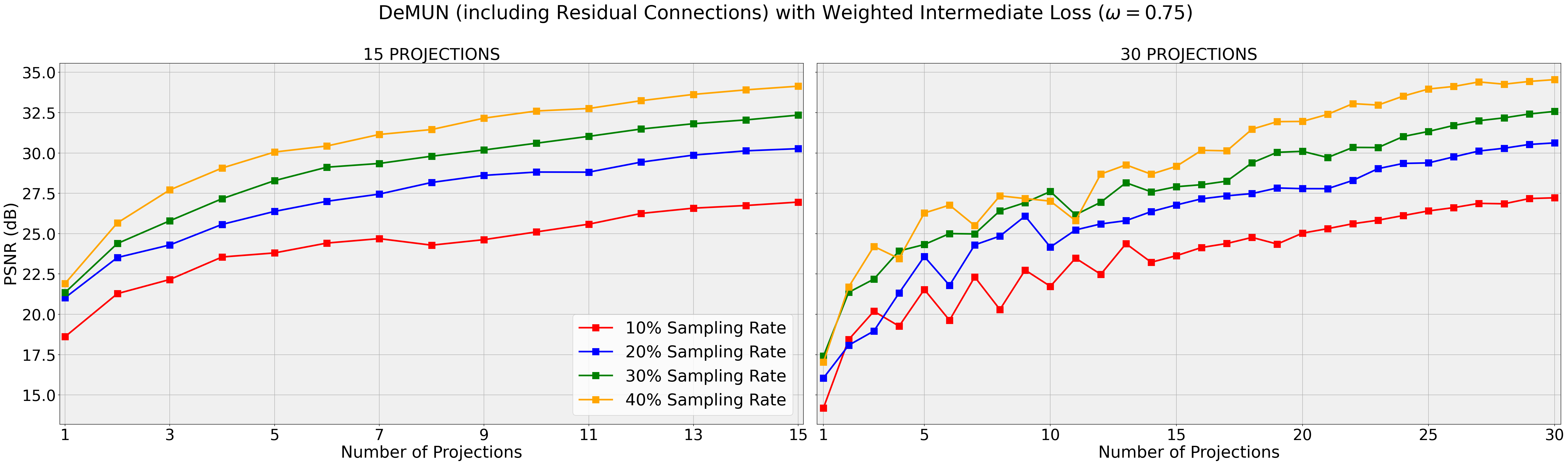}
\caption{DeMUN (including residual connections) with loss $\ell_{i, 0.75}$. The networks are trained for $T=15$ (left) and $T=30$ (right), and the graph displays the PSNR after each intermediate projection.}
\label{fig: DEMUN_R_SL_75}
\end{center}
\end{figure}

\begin{figure}[ht!]
\begin{center}
\includegraphics[width= 1.0\textwidth]{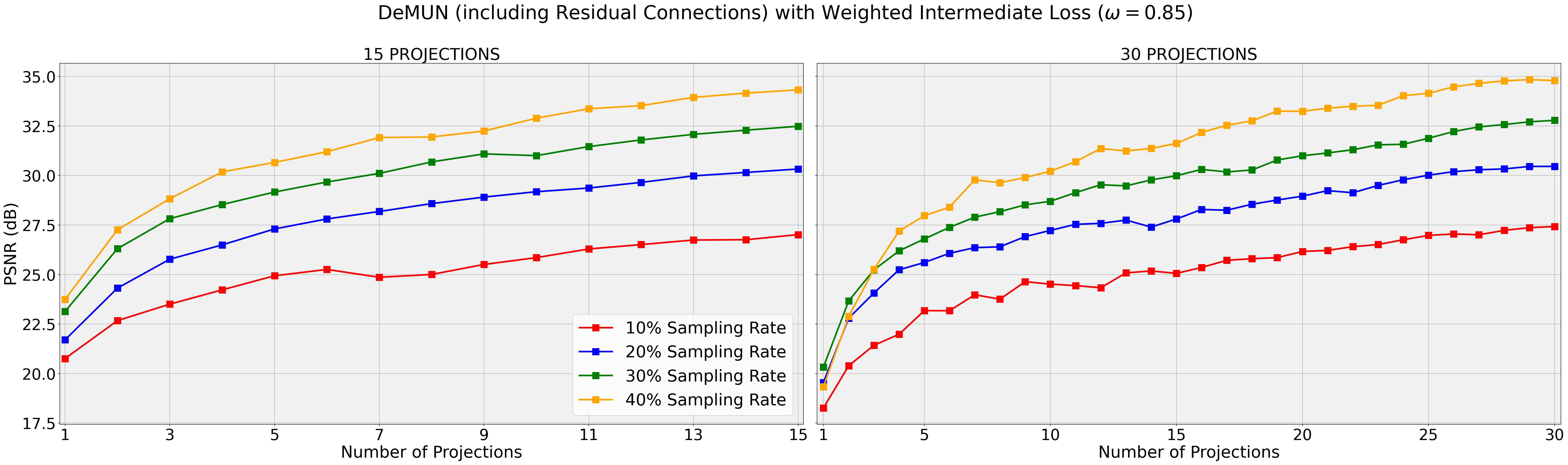}
\caption{DeMUN (including residual connections) with loss $\ell_{i, 0.85}$. The networks are trained for $T=15$ (left) and $T=30$ (right), and the graph displays the PSNR after each intermediate projection.}
\label{fig: DEMUN_R_SL_85}
\end{center}
\end{figure}

\begin{figure}[ht!]
\begin{center}
\includegraphics[width= 1.0\textwidth]{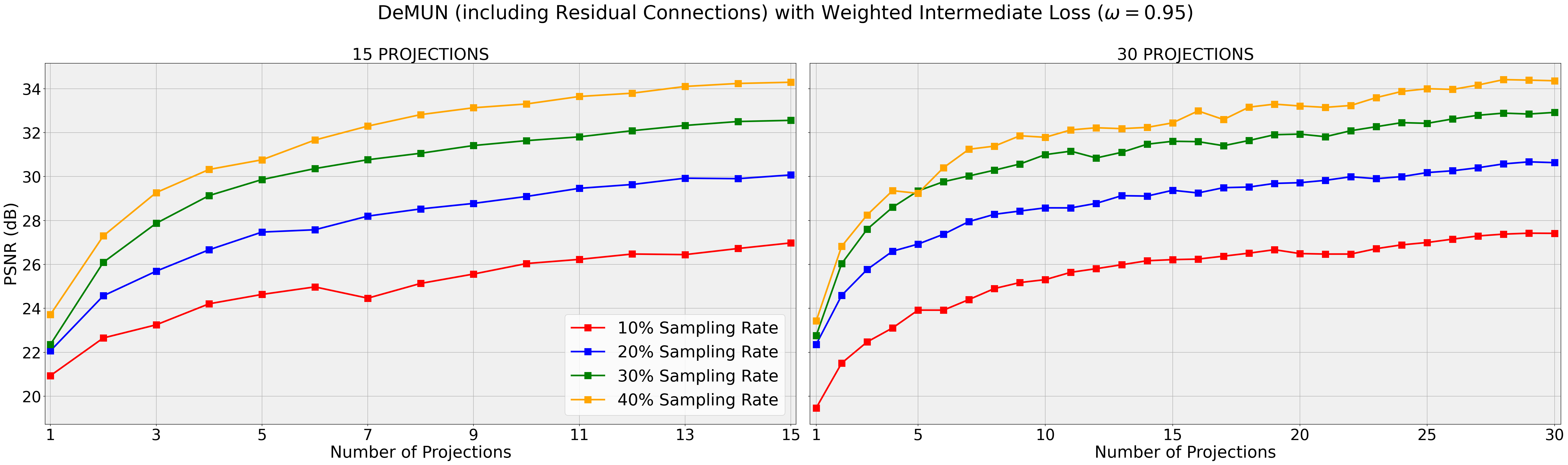}
\caption{DeMUN (including residual connections) with loss $\ell_{i, 0.95}$. The networks are trained for $T=15$ (left) and $T=30$ (right), and the graph displays the PSNR after each intermediate projection.}
\label{fig: DEMUN_R_SL_95}
\end{center}
\end{figure}

\begin{figure}[ht!]
\begin{center}
\includegraphics[width= 1.0\textwidth]{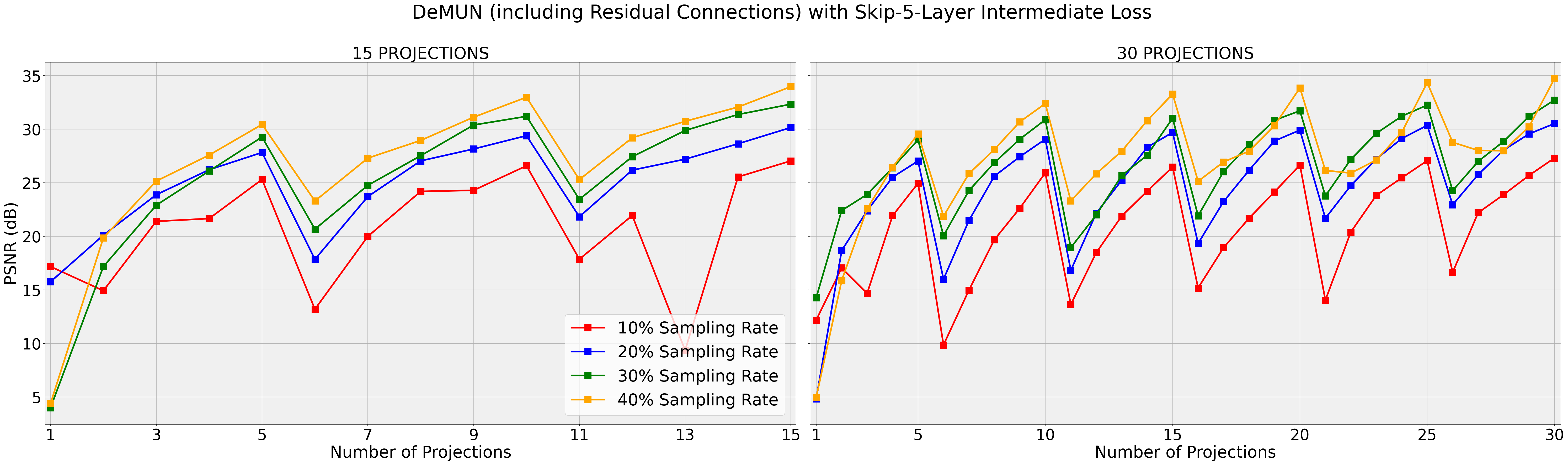}
\caption{DeMUN (including residual connections) with loss $\ell_{s, 5}$. The networks are trained for $T=15$ (left) and $T=30$ (right), and the graph displays the PSNR after each intermediate projection.}
\label{fig: DEMUN_R_Skip5}
\end{center}
\end{figure}

From the simulation results presented in Tables \ref{tb:diffloss:proj5}, \ref{tb:diffloss:proj15},  \ref{tb:diffloss:proj30} we are able to observe the following:
\begin{enumerate}
    \item \textbf{Minimal Impact for $\omega \ge 0.75$:} When $\omega \in \{1, 0.95, 0.85, 0.75\}$, the recovery performance remains relatively consistent, with negligible differences in the PSNR values.
    \item \textbf{Degradation with Small $\omega$:} For $\omega \in \{0.5, 0.25, 0.1, 0.01\}$, there is a noticeable decrease in reconstruction quality. This decline may be attributed to the exponential down-weighting of the initial layers, which causes the network to focus excessively on the later iterations, potentially leading to suboptimal convergence.
    \item \textbf{Skip-Layer Loss Performance:} We observe that the difference between the skip-5-layer loss and the unweighted intermediate loss decreases as the number of projection steps increases. When there are only 5 projection steps, $\ell_{s, 5}$ recovers the last-layer loss and $\ell_{i, 1}$ yields better performance. Under 30 projections, $\ell_{i, 1}$ is only better than $\ell_{s, 5}$ by a small margin. This indicates that the network can still achieve good recovery even when only supervising select layers. 
\end{enumerate}

These observations suggest that our network's performance is robust to moderate variations in the weighting of the loss function. So long as the intermediate outputs receive sufficient emphasis during training, the network can learn effective representations that lead to high-quality reconstructions. The decline in performance with small values of $\omega$ underscores the importance of adequately supervising the reconstruction of intermediate layers to guide the network toward the desirable recovery performance.

\subsection{Impact of the Complexity of the Projection Step} \label{ssec:complexityimpact}

The aim of this section is to examine Hypothesis 4, i.e., the effect of varying the capacity of the neural network projector $P_{\mathcal{C}}$, by changing the number of intermediate layers $L$ of the DnCNN architecture. We first assume that there is no additive measurement noise and consider $L = 3, 5, 10$, and $15$ layers. Our results are summarized in the following three tables. 

\begin{table}[ht!]
\small
        \centering
        \begin{tabular}{||c c c c c||} 
         \hline
         $m$ & $L = 3$ & $L = 5$ & $L = 10$ & $L = 15$\\ [0.5ex] 
         \hline\hline
         $0.1n$ & 26.33 & 26.29 & 26.28 & 25.99\\ 
         \hline
         $0.2n$ & 29.02 & 29.17 & 29.27 & 28.96\\
         \hline
         $0.3n$ & 30.71 & 30.95 & 31.16 & 30.87\\
         \hline
         $0.4n$ & 32.22 & 32.52 & 32.67 & 32.47\\
         \hline
        \end{tabular}
     \caption{Average PSNR (dB) for 2500 Test Images Under \textbf{5} Projection Steps}
\end{table}

\begin{table}[ht!]
\small
        \centering
        \begin{tabular}{||c c c c c||} 
         \hline
         $m$ & $L = 3$ & $L = 5$ & $L = 10$ & $L = 15$\\ [0.5ex] 
         \hline\hline
         $0.1n$ & 27.15 & 27.22 & 27.28 & 27.38 \\ 
         \hline
         $0.2n$ & 30.06 & 30.33 & 30.34 & 30.19 \\
         \hline
         $0.3n$ & 32.38 & 32.70 & 32.66 & 32.62 \\
         \hline
         $0.4n$ & 34.49 & 34.43 & 34.44 & 34.30 \\
         \hline
        \end{tabular}
     \caption{Average PSNR (dB) for 2500 Test Images Under \textbf{15} Projection Steps}
\end{table}

\begin{table}[ht!]
\small
        \centering
        \begin{tabular}{||c c c c c||} 
         \hline
         $m$ & $L = 3$ & $L = 5$ & $L = 10$ & $L = 15$ \\ [0.5ex] 
         \hline\hline
         $0.1n$ & 27.44 & 27.44 & 27.51 & 27.39 \\ 
         \hline
         $0.2n$ & 30.32 & 30.74 & 30.71 & 30.60 \\
         \hline
         $0.3n$ & 32.67 & 32.77 & 32.87 & 32.76 \\
         \hline
         $0.4n$ & 34.44 & 34.86 & 34.69 & 34.95 \\
         \hline
        \end{tabular}
 \caption{Average PSNR (dB) for 2500 Test Images Under \textbf{30} Projection Steps}
\end{table}

\FloatBarrier

We summarize our conclusions from the above three tables below:

\begin{enumerate}
\item Increasing the number of layers from \(5\) to \(15\) results in a negligible changes in the performance of DeMUNs, regardless of the number of projection steps. By comparing the above three tables, we observe that the number of projections has a significantly greater impact on performance than the number of layers within each projection.

\item By comparing $L=3$ and $L=5$, we conclude that reducing the depth too drastically $(L \leq 3)$ may impair the network's ability to learn complex features, as convolutional neural networks rely on multiple layers to capture hierarchical representations \cite{yamashita_convolutional_2018}.

\end{enumerate}

We acknowledge that the above conclusion may not necessarily extend to other projector architectures that do not rely on deep convolutional layers. Nevertheless, we believe this observation generalizes to other types of architectures when their capacity diminishes beyond a certain threshold, though we defer further investigation to future work. Additionally, these conclusions have yet to be validated for measurement matrices other than Gaussian ones and for scenarios where noise is present in the measurements. These cases will be addressed in Section \ref{subsection: projector capacity}.

\section{Robustness of DeMUNs}
\label{section: robustness}
In Section \ref{section: three hypothesis}, we have established, through extensive simulations, the superior performance of DeMUNs trained with unweighted intermediate loss $\ell_{i,1}$ and residual connections. The aim of this section is to assess the robustness of this configuration under various conditions. Specifically, we examine our network's performance under changes in the measurement matrix, the presence of additive noise, variations in input image resolution, and changes in projector capacity. These aspects represent the primary variables that practitioners must consider when deploying unrolled networks in real-world scenarios. Our extensive experiments demonstrate the adequacy and generalizability of our design choices. In the simulations presented in the following sections, we fix the image resolution to $50 \times 50$ when the resolution has not been specified.

\subsection{Robustness to the Sampling Matrix }\label{ssec:robustDCT}

We first investigate our network's performance under different sampling matrix structures. In addition to the Gaussian random matrix used previously, we consider a Discrete Cosine Transform (DCT) matrix of the form $A = SF \in \mathbb{R}^{m \times n}$ where $S \in \mathbb{R}^{m \times n}$ is an undersampling matrix and $F$ represents the 2D-DCT. We set the number of hidden layers for each projector (DnCNN) $L = 5$. Additional implementation details can be found in Appendix \ref{Experimental Setup}. There are a few points that we would like to clarify here:

\begin{enumerate}

\item Tabel \ref{table:forwardmodel_diversity} demonstrates that our network maintains excellent performance when considering DCT-type measurement matrices as well. The network effectively adapts to the DCT matrices, achieving comparable or better PSNR values than the Gaussian forward model. This suggests that our design choices made based on our simulations on Gaussian forward models offer good performance for other types of matrices as well.

\item The performance improvement DeMUNs gain from additional projection steps on DCT forward models is typically less than the improvement achieved with additional projections on Gaussian matrices. Again, we should emphasize that since overfitting does not seem to be an issue, as long as the recovery performance is concerned, the user does not need to worry about the number of projection steps when designing the network.

\end{enumerate}

\begin{table}[htbp]
    \centering
    \begin{tabular}{llccc}
        \toprule
        $m$ & Matrix& 5 Steps & 15 Steps & 30 Steps  \\ 
        \midrule
        \multirow{2}{*}{$0.1n$} 
            & Gaussian & 26.29 & 27.22 & 27.44 \\ 
            & DCT     & 28.42 & 28.47 & 28.53 \\
        \addlinespace
        \multirow{2}{*}{$0.2n$} 
            & Gaussian & 29.17 & 30.33 & 30.74 \\
            & DCT     & 30.00 & 30.37 & 30.48 \\
        \addlinespace
        \multirow{2}{*}{$0.3n$} 
            & Gaussian & 30.95 & 32.70 & 32.77 \\
            & DCT     & 31.50 & 32.15 & 32.19 \\
        \addlinespace
        \multirow{2}{*}{$0.4n$} 
            & Gaussian & 32.52 & 34.43 & 34.86 \\
            & DCT     & 33.21 & 33.90 & 34.03 \\
        \bottomrule
    \end{tabular}
    \caption{Average PSNR (dB) for 2500 Test Images Across Different Projection Steps}\label{table:forwardmodel_diversity}
\end{table}

\subsection{Robustness to Additive Noise}

Next, we introduce additive noise and obtain measurements of the form $y = Ax + w$ where $\omega \sim \mathcal{N}(0, \sigma^2 I)$. We want to see if our design choices still offer good performance in the presence of additive noise. The primary objective of this section is to demonstrate that the PSNR of DeMUN reconstructions gradually decreases as the noise level increases. Additionally, we aim to reaffirm that overfitting does not occur as the number of projections increases. Table \ref{table:SNR_measurements} shows the Signal-to-Noise Ratio (SNR)\footnote{The SNR for any image $x$ is calculated as $\|Ax\|_2^2/m\sigma^2$ given the sampling matrix $A$ and sampling rate $m$.}  of the measurements across different noise levels $\sigma$.

\begin{table}[htbp]
    \centering
    \begin{tabular}{llcccc}
        \toprule
        $m$ & Matrix & $\sigma = 0.01$ & $\sigma = 0.025$ & $\sigma = 0.05$  & $\sigma = 0.10$\\ 
        \midrule
        \multirow{2}{*}{$0.1n$} 
            & Gaussian & 32.19 & 24.23 & 18.21 & 12.19\\ 
            & DCT & 42.61 & 34.65 & 28.63 & 22.61\\
        \addlinespace
        \multirow{2}{*}{$0.2n$} 
            & Gaussian & 32.55 & 24.59 & 18.57 & 12.55\\
            & DCT & 39.61 & 31.65 & 25.63 & 19.61\\
        \addlinespace
        \multirow{2}{*}{$0.3n$} 
            & Gaussian & 32.57 & 24.62 & 18.59 & 12.57\\
            & DCT & 37.87 & 29.91 & 23.89 & 17.87\\
        \addlinespace
        \multirow{2}{*}{$0.4n$} 
            & Gaussian & 32.49 & 24.53 & 18.51 & 12.49\\
            & DCT & 36.63 & 28.67 & 22.65 & 16.63\\
        \bottomrule
    \end{tabular}
    \caption{Input SNR (dB) for Test Images Under Different Sampling Rates and Noise Levels.}
    \label{table:SNR_measurements}
\end{table}

\begin{table}[ht!]
\small
    \begin{subtable}[ht]{0.45\textwidth}
        \centering
        \begin{tabular}{||c c c c c||} 
         \hline
         $m \backslash \sigma$ & $0.01$ & $0.025$ & $0.05$ & $0.10$\\[0.5ex] 
         \hline\hline
         $0.1n$ & 26.14 & 25.38 & 24.51 & 22.85 \\ 
         \hline
         $0.2n$ & 28.76 & 27.83 & 26.36 & 24.41 \\ 
         \hline
         $0.3n$ & 30.43 & 29.29 & 27.61 & 25.39 \\ 
         \hline
         $0.4n$ & 31.80 & 30.34 & 28.46 & 26.07 \\ 
         \hline
        \end{tabular}
       \caption*{Gaussian Matrix}
    \end{subtable}
    \hfill
    \begin{subtable}[ht]{0.45\textwidth}
        \centering
        \begin{tabular}{||c c c c c||} 
         \hline
         $m \backslash \sigma$ & $0.01$ & $0.025$ & $0.05$ & $0.10$\\ [0.5ex] 
         \hline\hline
         $0.1n$ & 28.25 & 27.89 & 27.34 & 26.32 \\
         \hline
         $0.2n$ & 29.71 & 29.00 & 28.03 & 26.63 \\
         \hline
         $0.3n$ & 31.07 & 30.04 & 28.75 & 27.02 \\
         \hline
         $0.4n$ & 32.54 & 31.17 & 29.49 & 27.42 \\
         \hline
        \end{tabular}
        \caption*{DCT Matrix}
     \end{subtable}
     \caption{Average PSNR (dB) for 2500 Test Images Under \textbf{5} Projection Steps}\label{tabele1:noise:5proj}
\end{table}

\begin{table}[ht!]
\small
    \begin{subtable}[ht]{0.45\textwidth}
        \centering
        \begin{tabular}{||c c c c c||} 
         \hline
         $m \backslash \sigma$ & $0.01$ & $0.025$ & $0.05$ & $0.10$\\ [0.5ex] 
         \hline\hline
         $0.1n$ & 26.89 & 26.04 & 24.69 & 23.09 \\ 
         \hline
         $0.2n$ & 29.53 & 28.28 & 26.73 & 24.58 \\ 
         \hline
         $0.3n$ & 31.55 & 29.70 & 27.79 & 25.56 \\ 
         \hline
         $0.4n$ & 32.84 & 30.81 & 28.62 & 26.26 \\ 
         \hline
        \end{tabular}
       \caption*{Gaussian Matrix}
    \end{subtable}
    \hfill
    \begin{subtable}[ht]{0.45\textwidth}
        \centering
        \begin{tabular}{||c c c c c||} 
         \hline
         $m \backslash \sigma$ & $0.01$ & $0.025$ & $0.05$ & $0.10$\\[0.5ex] 
         \hline\hline
         $0.1n$ & 28.33 & 27.96 & 27.36 & 26.36 \\
         \hline
         $0.2n$ & 29.97 & 29.23 & 28.17 & 26.75 \\
         \hline
         $0.3n$ & 31.46 & 30.26 & 28.84 & 27.08 \\
         \hline
         $0.4n$ & 32.78 & 31.30 & 29.63 & 27.51 \\
         \hline
        \end{tabular}
        \caption*{DCT Matrix}
     \end{subtable}
     \caption{Average PSNR (dB) for 2500 Test Images Under \textbf{15} Projection Steps} \label{tabele2:noise:15proj}
\end{table}

\begin{table}[ht!]
\small
    \begin{subtable}[ht]{0.45\textwidth}
        \centering
        \begin{tabular}{||c c c c c||} 
         \hline
         $m \backslash \sigma$ & $0.01$ & $0.025$ & $0.05$ & $0.10$\\  [0.5ex] 
         \hline\hline
         $0.1n$ & 27.09 & 26.17 & 24.89 & 23.20 \\ 
         \hline
         $0.2n$ & 29.64 & 28.50 & 26.71 & 24.74 \\ 
         \hline
         $0.3n$ & 31.58 & 29.71 & 27.79 & 25.60 \\ 
         \hline
         $0.4n$ & 32.81 & 30.70 & 28.63 & 26.27 \\ 
         \hline
        \end{tabular}
       \caption*{Gaussian Matrix}
    \end{subtable}
    \hfill
    \begin{subtable}[ht]{0.45\textwidth}
        \centering
        \begin{tabular}{||c c c c c||} 
         \hline
         $m \backslash \sigma$ & $0.01$ & $0.025$ & $0.05$ & $0.10$\\  [0.5ex] 
         \hline\hline
         $0.1n$ & 28.34 & 27.97 & 27.36 & 26.35 \\
         \hline
         $0.2n$ & 30.05 & 29.16 & 28.21 & 26.74 \\
         \hline
         $0.3n$ & 31.43 & 30.30 & 28.91 & 27.14 \\
         \hline
         $0.4n$ & 32.90 & 31.33 & 29.58 & 27.56 \\
         \hline
        \end{tabular}
        \caption*{DCT Matrix}
     \end{subtable}
     \caption{Average PSNR (dB) for 2500 Test Images Under \textbf{30} Projection Steps}\label{tabele3:noise:30proj}
\end{table}

\FloatBarrier

We summarize some of our conclusions from Tables \ref{tabele1:noise:5proj}, \ref{tabele2:noise:15proj}, and \ref{tabele3:noise:30proj} below:

\begin{enumerate}
\item Although additive measurement noise predictably lowers the achievable PSNR, its impact on performance is relatively controlled. In particular, as the noise level increases, the PSNR degrades at a rate significantly slower than the corresponding decrease in the input SNR. This suggests that the network effectively suppresses the measurement noise. 

\item Our results indicate that as the noise level grows, the marginal benefit of additional projection steps diminishes. In other words, fewer projection steps often suffice to achieve comparable reconstruction quality. As mentioned before and is clear from the above tables, still increasing the number of projections does not hurt the reconstruction performance of the network. Hence, in scenarios where the noise level is not known, practitioners can choose a number that works well for the noiseless setting and use it for the noisy settings as well. 
\end{enumerate}

\subsection{Robustness of Hypothesis 4 to Sampling Matrix and Additive Noise}\label{subsection: projector capacity}

The main goal of this section is to evaluate the robustness of Hypothesis 4, particularly in response to changes in the measurement matrix and the presence of measurement noise. One might argue that at higher noise levels, using more convolutional layers would be beneficial for capturing more global information about the image and reducing noise. If this heuristic were valid, we would expect DeMUNs with larger \( L \) values to deliver better performance. However, as we will demonstrate below, this is not the case. The conclusions drawn in Section \ref{ssec:complexityimpact} hold true even in the presence of measurement noise and when the measurement matrix is non-Gaussian.

We first assume that there is no additive measurement noise and consider $L = 3, 5, 10$, and $15$ layers. We then evaluate the performance DeMUNs on DCT type matrices described in Section \ref{ssec:robustDCT}. To facilitate easy comparison, we also include the results reported in Section \ref{ssec:complexityimpact} for Gaussian matrices in the table below.

\begin{table}[ht!]
\small
    \begin{subtable}[ht]{0.48\textwidth}
        \centering
        \begin{tabular}{||c c c c c||} 
         \hline
         $m$ & $L = 3$ & $L = 5$ & $L = 10$ & $L = 15$\\ [0.5ex] 
         \hline\hline
         $0.1n$ & 26.33 & 26.29 & 26.28 & 25.99\\ 
         \hline
         $0.2n$ & 29.02 & 29.17 & 29.27 & 28.96\\
         \hline
         $0.3n$ & 30.71 & 30.95 & 31.16 & 30.87\\
         \hline
         $0.4n$ & 32.22 & 32.52 & 32.67 & 32.47\\
         \hline
        \end{tabular}
       \caption*{Gaussian Matrix}
    \end{subtable}
    \hfill
    \begin{subtable}[ht]{0.48\textwidth}
        \centering
        \begin{tabular}{||c c c c c||} 
         \hline
         $m$ & $L = 3$ & $L = 5$ & $L = 10$ & $L = 15$\\ [0.5ex] 
         \hline\hline
         $0.1n$ & 28.35 & 28.42 & 28.43 & 28.42\\ 
         \hline
         $0.2n$ & 29.94 & 30.00 & 30.01 & 29.96\\
         \hline
         $0.3n$ & 31.45 & 31.50 & 31.41 & 31.44\\
         \hline
         $0.4n$ & 33.06 & 33.21 & 33.23 & 33.20\\
         \hline
        \end{tabular}
        \caption*{DCT Matrix}
     \end{subtable}
     \caption{Average PSNR (dB) for 2500 Test Images Under \textbf{5} Projection Steps}
\end{table}

\begin{table}[ht!]
\small
    \begin{subtable}[ht]{0.48\textwidth}
        \centering
        \begin{tabular}{||c c c c c||} 
         \hline
         $m$ & $L = 3$ & $L = 5$ & $L = 10$ & $L = 15$\\ [0.5ex] 
         \hline\hline
         $0.1n$ & 27.15 & 27.22 & 27.28 & 27.38 \\ 
         \hline
         $0.2n$ & 30.06 & 30.33 & 30.34 & 30.19 \\
         \hline
         $0.3n$ & 32.38 & 32.70 & 32.66 & 32.62 \\
         \hline
         $0.4n$ & 34.49 & 34.43 & 34.44 & 34.30 \\
         \hline
        \end{tabular}
       \caption*{Gaussian Matrix}
    \end{subtable}
    \hfill
    \begin{subtable}[ht]{0.48\textwidth}
        \centering
        \begin{tabular}{||c c c c c||} 
         \hline
         $m$ & $L = 3$ & $L = 5$ & $L = 10$ & $L = 15$\\ [0.5ex] 
         \hline\hline
         $0.1n$ & 28.44 & 28.47 & 28.49 & 28.50 \\ 
         \hline
         $0.2n$ & 30.29 & 30.37 & 30.43 & 30.19 \\
         \hline
         $0.3n$ & 31.92 & 32.15 & 32.12 & 32.10 \\
         \hline
         $0.4n$ & 33.75 & 33.90 & 33.86 & 33.79 \\
         \hline
        \end{tabular}
        \caption*{DCT Matrix}
     \end{subtable}
     \caption{Average PSNR (dB) for 2500 Test Images Under \textbf{15} Projection Steps}
\end{table}

\begin{table}[ht!]
\small
    \begin{subtable}[ht]{0.48\textwidth}
        \centering
        \begin{tabular}{||c c c c c||} 
         \hline
         $m$ & $L = 3$ & $L = 5$ & $L = 10$ & $L = 15$ \\ [0.5ex] 
         \hline\hline
         $0.1n$ & 27.44 & 27.44 & 27.51 & 27.39 \\ 
         \hline
         $0.2n$ & 30.32 & 30.74 & 30.71 & 30.60 \\
         \hline
         $0.3n$ & 32.67 & 32.77 & 32.87 & 32.76 \\
         \hline
         $0.4n$ & 34.44 & 34.86 & 34.69 & 34.95 \\
         \hline
        \end{tabular}
       \caption*{Gaussian Matrix}
    \end{subtable}
    \hfill
    \begin{subtable}[ht]{0.48\textwidth}
        \centering
        \begin{tabular}{||c c c c c||} 
         \hline
         $m$ & $L = 3$ & $L = 5$ & $L = 10$ & $L = 15$ \\ [0.5ex] 
         \hline\hline
         $0.1n$ & 28.51 & 28.53 & 28.53 & 28.51 \\ 
         \hline
         $0.2n$ & 30.35 & 30.48 & 30.52 & 30.37 \\
         \hline
         $0.3n$ & 32.19 & 32.19 & 32.18 & 31.98 \\
         \hline
         $0.4n$ & 33.88 & 34.03 & 34.03 & 33.88 \\
         \hline
        \end{tabular}
        \caption*{DCT Matrix}
     \end{subtable}
     \caption{Average PSNR (dB) for 2500 Test Images Under \textbf{30} Projection Steps}
\end{table}

As evident from the above tables, the conclusion that increasing \( L \) from \( 5 \) to \( 15 \) does not provide a noticeable improvement remains valid for DCT-type matrices as well. Furthermore, one could argue that, in most cases for DCT-type matrices, the performance gain from increasing \( L \) from \( 3 \) to \( 5 \) is also marginal.

Next, we study the accuracy of Hypothesis 4 when additive noise is present in the measurements. Here, we consider three noise levels $\sigma \in \{0.01, 0.025, 0.05\}$ and test depths of $L = 3, 5$, and $10$.

\begin{table}[ht!]
\centering
\small
\begin{tabular}{||c|c c c|c c c|c c c||} 
\hline
\multirow{2}{*}{\begin{tabular}[c]{@{}c@{}}$m$\end{tabular}} & 
\multicolumn{3}{c|}{$\sigma = 0.01$} & 
\multicolumn{3}{c|}{$\sigma = 0.025$} & 
\multicolumn{3}{c||}{$\sigma = 0.05$} \\
& $L=3$ & $L=5$ & $L=10$ & $L=3$ & $L=5$ & $L=10$ & $L=3$ & $L=5$ & $L=10$ \\
\hline\hline
$0.1n$ & 26.05 & 26.14 & 26.07 & 25.44 & 25.38 & 25.50 & 24.35 & 24.51 & 24.45 \\
\hline
$0.2n$ & 28.69 & 28.76 & 28.79 & 27.66 & 27.83 & 27.86 & 26.25 & 26.36 & 26.40 \\
\hline
$0.3n$ & 30.38 & 30.43 & 30.51 & 29.14 & 29.29 & 29.40 & 27.44 & 27.61 & 27.67 \\
\hline
$0.4n$ & 31.61 & 31.80 & 31.90 & 30.24 & 30.34 & 30.50 & 28.33 & 28.46 & 28.50 \\
\hline
\end{tabular}
\caption{Average PSNR (dB) for 2500 Test Images Under \textbf{5} Projection Steps with Gaussian Matrix}
\end{table}

\begin{table}[ht!]
\centering
\small
\begin{tabular}{||c|c c c|c c c|c c c||} 
\hline
\multirow{2}{*}{\begin{tabular}[c]{@{}c@{}}$m$\end{tabular}} & 
\multicolumn{3}{c|}{$\sigma = 0.01$} & 
\multicolumn{3}{c|}{$\sigma = 0.025$} & 
\multicolumn{3}{c||}{$\sigma = 0.05$} \\
& $L=3$ & $L=5$ & $L=10$ & $L=3$ & $L=5$ & $L=10$ & $L=3$ & $L=5$ & $L=10$ \\
\hline\hline
$0.1n$ & 26.68 & 26.89 & 26.98 & 25.97 & 26.04 & 26.11 & 24.75 & 24.69 & 24.98 \\
\hline
$0.2n$ & 29.62 & 29.53 & 29.79 & 28.22 & 28.28 & 28.46 & 26.55 & 26.73 & 26.74 \\
\hline
$0.3n$ & 31.38 & 31.55 & 31.57 & 29.61 & 29.70 & 29.88 & 27.65 & 27.79 & 27.99 \\
\hline
$0.4n$ & 32.86 & 32.84 & 33.10 & 30.63 & 30.81 & 31.05 & 28.57 & 28.62 & 28.58 \\
\hline
\end{tabular}
\caption{Average PSNR (dB) for 2500 Test Images Under\textbf{ 15} Projection Steps with Gaussian Matrix}
\end{table}

\begin{table}[ht!]
\centering
\small
\begin{tabular}{||c|c c c|c c c|c c c||} 
\hline
\multirow{2}{*}{\begin{tabular}[c]{@{}c@{}}$m$\end{tabular}} & 
\multicolumn{3}{c|}{$\sigma = 0.01$} & 
\multicolumn{3}{c|}{$\sigma = 0.025$} & 
\multicolumn{3}{c||}{$\sigma = 0.05$} \\
& $L=3$ & $L=5$ & $L=10$ & $L=3$ & $L=5$ & $L=10$ & $L=3$ & $L=5$ & $L=10$ \\
\hline\hline
$0.1n$ & 26.95 & 27.09 & 27.03 & 26.03 & 26.17 & 26.19 & 24.81 & 24.89 & 25.04 \\
\hline
$0.2n$ & 29.53 & 29.64 & 30.06 & 28.31 & 28.50 & 28.59 & 26.59 & 26.71 & 26.93 \\
\hline
$0.3n$ & 31.42 & 31.58 & 31.83 & 29.61 & 29.71 & 29.98 & 27.83 & 27.79 & 28.11 \\
\hline
$0.4n$ & 32.81 & 32.81 & 33.06 & 30.70 & 30.70 & 30.94 & 28.49 & 28.63 & 28.72 \\
\hline
\end{tabular}
\caption{Average PSNR (dB) for 2500 Test Images Under \textbf{30} Projection Steps with Gaussian Matrix}
\end{table}

\begin{table}[ht!]
\centering
\small
\begin{tabular}{||c|c c c|c c c|c c c||} 
\hline
\multirow{2}{*}{\begin{tabular}[c]{@{}c@{}}$m$\end{tabular}} & 
\multicolumn{3}{c|}{$\sigma = 0.01$} & 
\multicolumn{3}{c|}{$\sigma = 0.025$} & 
\multicolumn{3}{c||}{$\sigma = 0.05$} \\
& $L=3$ & $L=5$ & $L=10$ & $L=3$ & $L=5$ & $L=10$ & $L=3$ & $L=5$ & $L=10$ \\
\hline\hline
$0.1n$ & 28.22 & 28.25 & 28.32 & 27.89 & 27.89 & 27.94 & 27.29 & 27.34 & 27.39 \\
\hline
$0.2n$ & 29.66 & 29.70 & 29.66 & 28.92 & 29.00 & 28.95 & 27.94 & 28.03 & 28.04 \\
\hline
$0.3n$ & 31.03 & 31.07 & 30.91 & 29.98 & 30.04 & 30.05 & 28.66 & 28.75 & 28.77 \\
\hline
$0.4n$ & 32.49 & 32.54 & 32.53 & 31.09 & 31.17 & 31.24 & 29.38 & 29.49 & 29.58 \\
\hline
\end{tabular}
\caption{Average PSNR (dB) for 2500 Test Images Under \textbf{5} Projection Steps with DCT Matrix}
\end{table}

\begin{table}[ht!]
\centering
\small
\begin{tabular}{||c|c c c|c c c|c c c||} 
\hline
\multirow{2}{*}{\begin{tabular}[c]{@{}c@{}}$m$\end{tabular}} & 
\multicolumn{3}{c|}{$\sigma = 0.01$} & 
\multicolumn{3}{c|}{$\sigma = 0.025$} & 
\multicolumn{3}{c||}{$\sigma = 0.05$} \\
& $L=3$ & $L=5$ & $L=10$ & $L=3$ & $L=5$ & $L=10$ & $L=3$ & $L=5$ & $L=10$ \\
\hline\hline
$0.1n$ & 28.30 & 28.33 & 28.37 & 27.93 & 27.96 & 28.02 & 27.30 & 27.36 & 27.36 \\
\hline
$0.2n$ & 29.88 & 29.97 & 30.01 & 29.12 & 29.23 & 29.28 & 28.08 & 28.17 & 28.26 \\
\hline
$0.3n$ & 31.26 & 31.46 & 31.48 & 30.14 & 30.26 & 30.38 & 28.79 & 28.84 & 28.93 \\
\hline
$0.4n$ & 32.67 & 32.78 & 32.90 & 31.19 & 31.30 & 31.41 & 29.49 & 29.63 & 29.67 \\
\hline
\end{tabular}
\caption{Average PSNR (dB) for 2500 Test Images Under \textbf{15} Projection Steps with DCT Matrix}
\end{table}

\begin{table}[ht!]
\centering
\small
\begin{tabular}{||c|c c c|c c c|c c c||} 
\hline
\multirow{2}{*}{\begin{tabular}[c]{@{}c@{}}$m$\end{tabular}} & 
\multicolumn{3}{c|}{$\sigma = 0.01$} & 
\multicolumn{3}{c|}{$\sigma = 0.025$} & 
\multicolumn{3}{c||}{$\sigma = 0.05$} \\
& $L=3$ & $L=5$ & $L=10$ & $L=3$ & $L=5$ & $L=10$ & $L=3$ & $L=5$ & $L=10$ \\
\hline\hline
$0.1n$ & 28.32 & 28.34 & 28.38 & 27.97 & 27.97 & 27.98 & 27.34 & 27.36 & 27.36 \\
\hline
$0.2n$ & 29.93 & 30.05 & 30.10 & 29.09 & 29.16 & 29.32 & 28.13 & 28.21 & 28.25 \\
\hline
$0.3n$ & 31.42 & 31.43 & 31.53 & 30.25 & 30.30 & 30.37 & 28.88 & 28.91 & 28.91 \\
\hline
$0.4n$ & 32.78 & 32.90 & 33.07 & 31.26 & 31.33 & 31.45 & 29.57 & 29.58 & 29.71 \\
\hline
\end{tabular}
\caption{Average PSNR (dB) for 2500 Test Images Under \textbf{30} Projection Steps with DCT Matrix}
\end{table}

\FloatBarrier
These results strongly suggest that even in the presence of additive noise, increasing $L$ does not offer substantial gain in the performance of DeMUNs. Given that the improvement in recovery performance is marginal when increasing projector capacity, this suggests that simple architectures like DnCNN with very few convolutional layers may be sufficient for practical applications where measurement noise is present, offering potential computational savings without significant performance degradation. 

\FloatBarrier
\subsection{Robustness to Image Resolution}
\label{subsection: image resolution}

Finally, we assess DeMUN's performance across different image resolutions. We test resolutions of $32 \times 32, 50 \times 50$, $64 \times 64$, and $80 \times 80$ fixing the measurement matrices and removing measurement noise. There are two main questions we aim to address here: (1) Do we need more or fewer projections as we increase the number of projections? (2) How should we set the number of layers $L$ in the projection as we increase/decrease resolution?

We first set the number of intermediate layers of each projector to $L=5$, and run the simulations for different number of projections. 

\begin{table}[ht!]
\small
    \begin{subtable}[ht]{0.48\textwidth}
        \centering
        \begin{tabular}{||c c c c c||} 
         \hline
         $m$ & 32 $\times$ 32 & 50 $\times$ 50 & 64 $\times$ 64 & 80 $\times$ 80\\ [0.5ex] 
         \hline\hline
         $0.1n$ & 26.02 & 26.29 & 26.79 & 27.20 \\ 
         \hline
         $0.2n$ & 28.41 & 29.17 & 29.51 & 29.93 \\
         \hline
         $0.3n$ & 30.63 & 30.95 & 31.48 & 31.92 \\
         \hline
         $0.4n$ & 32.04 & 32.52 & 32.96 & 33.31 \\
         \hline
        \end{tabular}
       \caption*{Gaussian Matrix}
    \end{subtable}
    \hfill
    \begin{subtable}[ht]{0.48\textwidth}
        \centering
        \begin{tabular}{||c c c c c||} 
         \hline
         $m$ & 32 $\times$ 32 & 50 $\times$ 50 & 64 $\times$ 64 & 80 $\times$ 80\\ [0.5ex] 
         \hline\hline
         $0.1n$ & 28.61 & 28.42 & 28.53 & 28.66 \\ 
         \hline
         $0.2n$ & 30.16 & 30.00 & 30.43 & 30.57 \\
         \hline
         $0.3n$ & 31.08 & 31.50 & 31.97 & 32.40 \\
         \hline
         $0.4n$ & 32.83 & 33.21 & 33.51 & 33.97 \\
         \hline
        \end{tabular}
        \caption*{DCT Matrix}
     \end{subtable}
     \caption{Average PSNR (dB) for 2500 Test Images Under \textbf{5} Projection Steps}
\end{table}

\begin{table}[ht!]
\small
    \begin{subtable}[ht]{0.48\textwidth}
        \centering
        \begin{tabular}{||c c c c c||} 
         \hline
         $m$ & 32 $\times$ 32 & 50 $\times$ 50 & 64 $\times$ 64 & 80 $\times$ 80\\ [0.5ex] 
         \hline\hline
         $0.1n$ & 26.86 & 27.22 & 27.65 & 27.97 \\ 
         \hline
         $0.2n$ & 29.74 & 30.33 & 30.74 & 31.30 \\
         \hline
         $0.3n$ & 31.76 & 32.70 & 33.01 & 33.25 \\
         \hline
         $0.4n$ & 33.68 & 34.43 & 34.89 & 35.50 \\
         \hline
        \end{tabular}
       \caption*{Gaussian Matrix}
    \end{subtable}
    \hfill
    \begin{subtable}[ht]{0.48\textwidth}
        \centering
        \begin{tabular}{||c c c c c||} 
         \hline
         $m$ & 32 $\times$ 32 & 50 $\times$ 50 & 64 $\times$ 64 & 80 $\times$ 80\\ [0.5ex] 
         \hline\hline
         $0.1n$ & 28.74 & 28.47 & 28.62 & 28.78 \\ 
         \hline
         $0.2n$ & 30.48 & 30.37 & 30.88 & 31.09 \\
         \hline
         $0.3n$ & 31.66 & 32.15 & 32.36 & 33.18 \\
         \hline
         $0.4n$ & 33.36 & 33.90 & 34.32 & 34.91 \\
         \hline
        \end{tabular}
        \caption*{DCT Matrix}
     \end{subtable}
     \caption{Average PSNR (dB) for 2500 Test Images Under \textbf{15} Projection Steps}
\end{table}

\begin{table}[ht!]
\small
    \begin{subtable}[ht]{0.48\textwidth}
        \centering
        \begin{tabular}{||c c c c c||} 
         \hline
         $m$ & 32 $\times$ 32 & 50 $\times$ 50 & 64 $\times$ 64 & 80 $\times$ 80\\ [0.5ex] 
         \hline\hline
         $0.1n$ & 27.20 & 27.44 & 28.18 & 28.31 \\ 
         \hline
         $0.2n$ & 29.78 & 30.74 & 30.90 & 31.54 \\
         \hline
         $0.3n$ & 32.26 & 32.77 & 33.22 & 33.86 \\
         \hline
         $0.4n$ & 33.70 & 34.86 & 34.99 & 35.54 \\
         \hline
        \end{tabular}
       \caption*{Gaussian Matrix}
    \end{subtable}
    \hfill
    \begin{subtable}[ht]{0.48\textwidth}
        \centering
        \begin{tabular}{||c c c c c||} 
         \hline
         $m$ & 32 $\times$ 32 & 50 $\times$ 50 & 64 $\times$ 64 & 80 $\times$ 80\\ [0.5ex] 
         \hline\hline
         $0.1n$ & 28.70 & 28.53 & 28.64 & 28.83 \\ 
         \hline
         $0.2n$ & 30.49 & 30.48 & 30.86 & 31.15 \\
         \hline
         $0.3n$ & 31.62 & 32.19 & 32.58 & 33.35 \\
         \hline
         $0.4n$ & 33.46 & 34.03 & 34.36 & 35.02 \\
         \hline
        \end{tabular}
        \caption*{DCT Matrix}
     \end{subtable}
     \caption{Average PSNR (dB) for 2500 Test Images Under \textbf{30} Projection Steps}
\end{table}

\FloatBarrier

We observe that as the image resolution increases, the network's recovery performance generally improves (for fixed number of projections). This is likely due to the presence of more information in higher-resolution images, which aids the network in learning more detailed structural properties. As observed previously, increasing the number of projections continues to enhance the performance of DeMUNs when trained with the intermediate loss function (\(\omega = 1\)) and the residual connection. However, this improvement eventually plateaus, and the incremental gains become negligible beyond a certain point. For example, in nearly all the simulations presented above, the performance improvement from increasing the number of projections from \(15\) to \(30\) is minimal and may not justify the additional computational burden.  

Our next set of simulations aims to investigate the impact of the number of layers in the projection step (implemented using DnCNN) on the performance of DeMUNs as the resolution increases. Under the same setting as above, we increase the neural network projector capacity by setting the number of intermediate layers $L = 10$ and check if our conclusions remain valid. 

\begin{table}[ht!]
\small
    \begin{subtable}[ht]{0.48\textwidth}
        \centering
        \begin{tabular}{||c c c c c||} 
         \hline
         $m$ & 32 $\times$ 32 & 50 $\times$ 50 & 64 $\times$ 64 & 80 $\times$ 80\\ [0.5ex] 
         \hline\hline
         $0.1n$ & 25.70 & 26.28 & 26.58 & 27.28 \\ 
         \hline
         $0.2n$ & 28.52 & 29.27 & 29.65 & 30.21 \\
         \hline
         $0.3n$ & 30.40 & 31.16 & 31.75 & 32.07 \\
         \hline
         $0.4n$ & 31.95 & 32.67 & 33.25 & 33.69 \\
         \hline
        \end{tabular}
       \caption*{Gaussian Matrix}
    \end{subtable}
    \hfill
    \begin{subtable}[ht]{0.48\textwidth}
        \centering
        \begin{tabular}{||c c c c c||} 
         \hline
         $m$ & 32 $\times$ 32 & 50 $\times$ 50 & 64 $\times$ 64 & 80 $\times$ 80\\ [0.5ex] 
         \hline\hline
         $0.1n$ & 28.69 & 28.43 & 28.57 & 28.74 \\ 
         \hline
         $0.2n$ & 29.80 & 30.01 & 30.50 & 30.63 \\
         \hline
         $0.3n$ & 31.06 & 31.41 & 31.97 & 32.46 \\
         \hline
         $0.4n$ & 32.64 & 33.23 & 33.70 & 34.18 \\
         \hline
        \end{tabular}
        \caption*{DCT Matrix}
     \end{subtable}
     \caption{Average PSNR (dB) for 2500 Test Images Under \textbf{5} Projection Steps}
\end{table}

\begin{table}[ht!]
\small
    \begin{subtable}[ht]{0.48\textwidth}
        \centering
        \begin{tabular}{||c c c c c||} 
         \hline
         $m$ & 32 $\times$ 32 & 50 $\times$ 50 & 64 $\times$ 64 & 80 $\times$ 80\\ [0.5ex] 
         \hline\hline
         $0.1n$ & 26.65 & 27.28 & 27.94 & 28.24 \\ 
         \hline
         $0.2n$ & 29.68 & 30.34 & 30.79 & 31.28 \\
         \hline
         $0.3n$ & 32.05 & 32.66 & 33.11 & 33.62 \\
         \hline
         $0.4n$ & 33.51 & 34.44 & 34.89 & 35.50 \\
         \hline
        \end{tabular}
       \caption*{Gaussian Matrix}
    \end{subtable}
    \hfill
    \begin{subtable}[ht]{0.48\textwidth}
        \centering
        \begin{tabular}{||c c c c c||} 
         \hline
         $m$ & 32 $\times$ 32 & 50 $\times$ 50 & 64 $\times$ 64 & 80 $\times$ 80\\ [0.5ex] 
         \hline\hline
         $0.1n$ & 28.73 & 28.49 & 28.63 & 28.84 \\ 
         \hline
         $0.2n$ & 30.30 & 30.43 & 30.75 & 31.10 \\
         \hline
         $0.3n$ & 31.38 & 32.12 & 32.55 & 33.19 \\
         \hline
         $0.4n$ & 33.23 & 33.86 & 34.36 & 34.80 \\
         \hline
        \end{tabular}
        \caption*{DCT Matrix}
     \end{subtable}
     \caption{Average PSNR (dB) for 2500 Test Images Under \textbf{15} Projection Steps}
\end{table}

\begin{table}[ht!]
\small
    \begin{subtable}[ht]{0.48\textwidth}
        \centering
        \begin{tabular}{||c c c c c||} 
         \hline
         $m$ & 32 $\times$ 32 & 50 $\times$ 50 & 64 $\times$ 64 & 80 $\times$ 80\\ [0.5ex] 
         \hline\hline
         $0.1n$ & 27.10 & 27.51 & 28.13 & 28.67 \\ 
         \hline
         $0.2n$ & 30.31 & 30.71 & 31.21 & 31.50 \\
         \hline
         $0.3n$ & 32.09 & 32.87 & 33.40 & 34.12 \\
         \hline
         $0.4n$ & 33.92 & 34.69 & 35.40 & 35.94 \\
         \hline
        \end{tabular}
       \caption*{Gaussian Matrix}
    \end{subtable}
    \hfill
    \begin{subtable}[ht]{0.48\textwidth}
        \centering
        \begin{tabular}{||c c c c c||} 
         \hline
         $m$ & 32 $\times$ 32 & 50 $\times$ 50 & 64 $\times$ 64 & 80 $\times$ 80\\ [0.5ex] 
         \hline\hline
         $0.1n$ & 28.61 & 28.53 & 28.68 & 28.90 \\ 
         \hline
         $0.2n$ & 30.31 & 30.52 & 30.97 & 31.27 \\
         \hline
         $0.3n$ & 31.58 & 32.18 & 32.65 & 33.36 \\
         \hline
         $0.4n$ & 33.21 & 34.03 & 34.51 & 35.15 \\
         \hline
        \end{tabular}
        \caption*{DCT Matrix}
     \end{subtable}
     \caption{Average PSNR (dB) for 2500 Test Images Under \textbf{30} Projection Steps}
\end{table}

\FloatBarrier

These results confirm Hypothesis 4 again. Even as we increase the resolution of the images, the impact of the number of layers from $5$ to $10$ on the recovery performance remains negligible. 

Before concluding this section, we would like to highlight another key insight from the simulations above. Notably, a significant improvement in the reconstruction accuracy of DeMUNs is achieved as the resolution increases. This observation suggests an intriguing direction for future research: reducing the computational complexity of training DeMUNs to enable their application to higher-resolution images, thereby achieving even greater accuracy.

\section{Conclusion}

In this paper, we conducted a comprehensive empirical study on the design choices for unrolled networks in solving linear inverse problems. As our first step, we introduced the Deep Memory Unrolled Network (DeMUN), that leverages the history of all gradients and generalizes a wide range of existing unrolled networks. This approach was designed to (1) let the data decide on the optimal choice of algorithm to be unrolled, and hence (2) improve recovery performance. A byproduct of our choice is that the users do not need to decide on the choice of the algorithm they need to unroll. 

Through extensive simulations, we demonstrated that training DeMUN with an unweighted intermediate loss function and incorporating residual connections represents the best existing practice (among the ones studied in this paper) for optimizing these networks. This approach delivers superior performance compared to existing unrolled algorithms, highlighting its effectiveness and versatility.

We also presented experiments that exhibit the robustness of our design choices to a wide range of conditions, including different measurement matrices, additive noise levels, and image resolutions. Hence, our results offer practical guidelines and rules of thumb for selecting the loss function for training, structuring the unrolled network, determining the required number of projections, and deciding on the appropriate number of layers. These insights simplify the design and optimization of such networks for a wide range of applications, and we expect them to serve as a useful reference for researchers and practitioners in designing effective unrolled networks for linear inverse problems across various settings.

\bibliographystyle{unsrtnat}
\bibliography{references}

\appendix

\section{Experimental Setup}
\label{Experimental Setup}

Below, we discuss the implementation details abbreviated in the sections above. Our networks are trained with NVIDIA A100 SXM4, H100 PCIE, H100 SXM5, and GH200.\footnote{The accompanying code can be found at \url{https://github.com/YuxiChen25/Memory-Net-Inverse}.}

\subsection{Implementation Details}
Our dataset is generated from the 50K validation images from the 2012 ImageNet Large Scale Visual Recognition Challenge (ILSVRC2012).\footnote{ \url{https://www.image-net.org/challenges/LSVRC/2012/}} For each image from ILSVRC2012, we first convert it to grayscale and then crop the center $3k \times 3k$ region where $k = 32, 50, $ $64$, or $80$ depending on the image resolution. Then, each cropped image is converted into 9 images of size $k \times k$ where each smaller image is individually converted into a length $n = k^2$ vector $\bar{x}$.  

To generate the observation matrix $A \in \mathbb{R}^{m \times n}$ for Gaussian matrices, we draw $A_{ij} \sim \mathcal{N}(0, 1/m)$. For DCT matrices, the measurement matrix takes the form $A = SF$ where $S \in \mathbb{R}^{m \times n}$ is an under-sampling matrix and $F \in \mathbb{R}^{n \times n}$ is the 2D-Discrete Cosine Transform matrix generated by the Kronecker product of two 1D-DCT matrices of size $k \times k$. To generate the subsampling matrix $S$, we adopt the following policy: for 10\% sampling rate, we always sample the $10\%$ of the 2D-DCT transformed image located approximately in the top-left corner by fixing the indices in advance. For each increased sampling rate, we randomly sample indices located elsewhere in the transformed image. For both settings, we normalize the measurement matrix as outlined in Section \ref{subsection: initialization} to obtain $A'$. Then, each observation is generated through the process $y = A'x + w$ where $x$ is the normalized vector $x = \bar{x} / 255$. This generates 450K compressed measurements $y \in \mathbb{R}^n$ along with their ground-truth values $x$.

\subsection{Training and Evaluation Details}

\subsubsection{Training and Test Data}
To train and test the network, we take the first 25K images of the processed dataset and allot $2500$ images for testing and $22500$ for training. The training set is further partitioned into 18K images for training the network and $4500$ images for validation. For the training of the unrolled networks, we use a batch size of $32$ and set the number of training epochs to 300. The learning rate is set to 1e-4 using the ADAM Optimizer without any regularization. We select the model weights corresponding to the epoch that has the lowest validation loss evaluated on the $4500$ images using the mean-squared-error. The remaining 2500 images are used to report the test PSNR above.

\subsubsection{Initialization}
\label{subsection: initialization}
In the case of Deep-memory unrolled networks (DeMUNs), the trainable weights can be partitioned into two categories: (1) the weights of neural networks that are used in the projector units of Figure \ref{fig:unrollednetworkdiagram}, and (2) the weights $\alpha^i$ and $\{\beta_j^i\}$ in the $1 \times 1$ convolution that are used in the loss reducer blocks. We adopt the following approach to initialize the weights:
\begin{enumerate}
\item We calculate the maximum $\ell^2$ row norm of the matrix: $\|A\|_{\infty, 2} = \max_{1 \leq i \leq m}\sqrt{\sum_{j = 1}^{n}a_{ij}^2}$. Subsequently, we normalize $A$ according to $A' = A / \|A\|_{\infty, 2}$ to obtain our sampling matrix. 

\item For each loss reducer $i$, we initialize the first two terms $\alpha^i, \beta_i^i$ to $1$ and set all other weights to $0$.

\item All other network projector weights $\{P_{\mathcal{C}}^t\}_t$ are initialized randomly \cite{mousavi_learning_2017, zhang_ista-net_2018, zhang_amp-net_2021}. For instance in PyTorch, convolutional weights are initialized by sampling from Uniform$(-\sqrt{k}, -\sqrt{k})$, where \[k = \dfrac{groups}{C_{in} \cdot \prod_{i = 0}^1 kernel size_i}\] Here, $groups$ specifies independent channel groups (defaulting to 1), $C_{in}$ is the number of input channels, and $kernel size_i$ represents the kernel size along the \textit{i}-th dimension.\footnote{\url{https://pytorch.org/docs/stable/generated/torch.nn.Conv2d.html}}

\end{enumerate}

The normalization applied in Step 1 ensures that our training process is robust to the scaling of the measurement matrix (or forward operator) \( A \). Note that multiplying the measurement matrix by a factor, such as \( 0.001 \) or \( 1000 \), does not affect the inherent complexity of the problem. Hence, we expect our recovery method to produce the same estimate. However, in neural network models, where numerous multiplications and additions occur, extremely large or small values can lead to numerical issues. Additionally, due to the non-convexity of the training error, an initialization that performs well at one scale may not perform as effectively at another scale of the measurement matrix, since it may lead to a different local minima. The normalization introduced in Step 1 is designed to address these challenges, ensuring that both the initialization and the performance of the learned models remain robust to the scaling of \( A \).

For Step 2, we note that the first two memory terms $\alpha^i$ and $\beta_i^i$ correspond to $x^i$ and $A^T(y - Ax^i)$. Initializing these terms with $1$ recovers the standard projected gradient descent step $\tilde{x}^i = x^i + A^T(y - Ax^i)$ with unit step size to be fed into the neural network projector. Therefore, this scheme can be seen as initializing the memory terms to be standard projected gradient descent and making the weights corresponding to  terms $\{\beta_j^i\}_{j \in \{0, \dots, i - 1\}} $ in the memory to be learned as the training process progresses.

\subsection{Unrolled Algorithms}
\label{appendix: unrolled algs}

Here, we provide supplementary details on the unrolled networks based on Nesterov's First-Order method and Approximate Message passing used as comparison methods in Section \ref{subsection: Hypothesis 1}.

\subsubsection{Unrolled Nesterov's First-Order Method}

For Nesterov's first-order method, each iteration proceeds as follows starting from $i = 0$:

\[\tilde{x}^i = x_n^i + \mu A^T(y - Ax_n ^i)\]
\[x^{i + 1} = P_{\mathcal{C}}(\tilde{x}^i)\]
\[x_n^{i+1} = x^{i+1} + \bigg(\dfrac{t_{i+1} - 1}{t_{i + 2}}\bigg)(x^{i+1} - x^{i})\]

Here, $x_n^0 = x^0 = 0$ and $t_{i + 1} = \dfrac{1 + \sqrt{1 + 4t_i^2}}{2}$ with $t_1 = 1$. The weights multiplying the linear combination are fixed and not backpropagated during the optimization process.

\subsubsection{Unrolled Approximate Message Passing}

For unrolled Approximate Message Passing, each iteration proceeds in the following way:

\[\tilde{x}^i = x^i + A^T z^i\]
\[x^{i + 1} = P_{\mathcal{C}}(\tilde{x}^i)\]
\[z^{i + 1} = y - Ax^{i + 1} + z^i \text{div}(P_{\mathcal{C}}(x^i + A^T z^i)) / m\]

Here, $z^0 = y$ and $\text{div}(\cdot)$ is an estimate of the divergence of the projector estimated using a single Monte-Carlo sample \cite{metzler_denoising_2016}. 

\end{document}